\documentclass[fleqn,usenatbib]{mnras}

\usepackage{newtxtext,newtxmath}

\usepackage[T1]{fontenc}
\usepackage{ae,aecompl}

\usepackage{graphicx}	
\usepackage{amsmath}	
\usepackage{amssymb}	
\usepackage{aas_macros}
\usepackage{upgreek}
\usepackage{color}
\usepackage{pdflscape}

\newcommand{\err}[2]{$\substack{+#1 \\ -#2}$} 
\newcommand{\rerr}[2]{$\substack{+#2 \\ -#1}$}
\newcommand{\gp}{\phantom{--}}                
\newcommand\fdwarf{\hbox{f$_{\rm dwarf}$}\,}
\newcommand\umicron{\,\hbox{$\upmu$m}}

\title[KINETyS II: K-band extension]{KINETyS II: Constraints on spatial variations of the stellar initial mass function from K-band spectroscopy \thanks{Based on observations obtained at the Very Large Telescope of the European Southern Observatory. Programme ID: 097.B-0882(A)}}

\author[P. D. Alton et al.]{
	P. D. Alton,$^{1}$\thanks{E-mail: padraig.alton@durham.ac.uk}
	R. J. Smith,$^{1}$
	and J. R. Lucey,$^{1}$
	\\
	$^{1}$Centre for Extragalactic Astronomy, Department of Physics, Durham University, South Road, Durham DH1 3LE, UK \\
}

\date{Accepted 2018 May 4th. Received 2018 April 7; in original form 2017 July 12.}

\pubyear{2018}

\begin{document}
	\label{firstpage}
	\pagerange{\pageref{firstpage}--\pageref{lastpage}}
	\maketitle
	
\begin{abstract}
		
We investigate the spatially resolved stellar populations of a sample of seven nearby massive Early-type galaxies (ETGs), using optical and near infrared data, including K-band spectroscopy. This data offers good prospects for mitigating the uncertainties inherent in stellar population modelling by making a wide variety of strong spectroscopic features available. We report new VLT-KMOS measurements of the average empirical radial gradients out to the effective radius in the strengths of the Ca\,I 1.98\umicron\, and 2.26\umicron\, features, the Na\,I 2.21\umicron\, line, and the CO 2.30\umicron\, bandhead. Following previous work, which has indicated an excess of dwarf stars in the cores of massive ETGs, we pay specific attention to radial variations in the stellar initial mass function (IMF) as well as modelling the chemical abundance patterns and stellar population ages in our sample. Using state-of-the-art stellar population models we infer an [Fe/H] gradient of --0.16$\pm$0.05 per dex in fractional radius and an average [Na/Fe] gradient of --0.35$\pm$0.09. We find a large but radially-constant enhancement to [Mg/Fe] of $\sim$\,0.4 and a much lower [Ca/Fe] enhancement of $\sim$\,0.1. Finally, we find no significant IMF radial gradient in our sample on average and find that most galaxies in our sample are consistent with having a Milky Way-like IMF, or at most a modestly bottom heavy IMF (e.g. less dwarf enriched than a single power law IMF with the Salpeter slope).
		
\end{abstract}
	
	
	
	
\section{Introduction}
	
When studying an evolved population of stars, one of the key assumptions made concerns the form of the stellar initial mass function (IMF). The IMF -- the initial distribution of stellar masses in a newly-formed stellar population -- determines the future chemical evolution of the host galaxy, the rate of supernova explosions, and a variety of key observables. These include, critically, the mass-to-light ratio of the stellar population; any calculation of the stellar mass contained in a galaxy therefore depends on the IMF.
	
In our own galaxy, methods based on direct counting of stars have revealed that the IMF is remarkably uniform despite variations in environmental conditions (e.g. as reviewed by \citealp{2010ARA&A..48..339B}) with few exceptions (see e.g. \citealp{2003ApJ...590..348L}, \citealp{2012MNRAS.422.2246M}). Above 1M$_{\odot}$ the IMF is well-described by a power law \citep{1955ApJ...121..161S} and at lower masses this power law turns over (see \citealp{2001MNRAS.322..231K}; \citealp{2003PASP..115..763C}). Nevertheless, it is unclear whether the IMF in other galaxies has a similar functional form.
	
In unresolved stellar populations constraining the IMF is challenging, but attempts to do so have met with a certain amount of success in recent years. A key result has been the measurement, using multiple different techniques, of IMF variations in massive early-type galaxies (ETGs). In these galaxies the IMF appears to be `bottom-heavy', meaning that their stellar populations contain a larger fraction of dwarf stars than those in the Milky-Way: this can increase the mass-to-light (M/L) ratio of these galaxies by as much as a factor of two. 

Some methods yielding bottom-heavy IMFs in ETGs arrive at the result indirectly, via constraints on the M/L ratio from gravitational lensing (e.g. \citealp{2010ApJ...721L.163A}, but see also \citealp{2013MNRAS.434.1964S}) or dynamics (e.g. \citealp{2012Natur.484..485C}), or both \citep{2011MNRAS.415..545T}. There are alternative interpretations of enhanced M/L ratios: variations in the high-mass end of the IMF can also increase the M/L of an old stellar population due to the inclusion of more dim stellar remnants and variations in the dark-matter content of a galaxy can also play a role in some cases.

An alternative, spectroscopic class of methods provide direct support for the bottom-heavy interpretation. Such methods involve fitting models to galaxy spectra. Certain spectral features are strong/weak in the spectra of dwarf/giant stars (or vice-versa) with the same surface temperature, chemical composition, and age. In old stellar populations the strength of these features therefore provides information about the relative contribution to the total light from dwarf and giant stars. 

These methods have been used by, amongst others, \cite{2012ApJ...760...71C} (hereafter CvD12b), \cite{2012MNRAS.426.2994S}, \cite{2012ApJ...753L..32S}, \cite{2013MNRAS.433.3017L}, \cite{2015MNRAS.447.1033M}, and \cite{2017ApJ...841...68V} to infer bottom-heavy IMFs in the most massive ETGs, although this result has been challenged by e.g. \cite{2015MNRAS.452..597Z}, \cite{2016ApJ...821...39M}, and \cite{2016arXiv161200364V}.
	
Spectroscopic models suffer from degeneracies between different properties, e.g. a feature sensitive to the IMF will also be sensitive to the abundance of one or more chemical elements. Simultaneous consideration of a wide variety of spectral features with different sensitivities to the underlying properties of the stellar population can break these degeneracies. Measurements of near-infrared features have begun to play a larger role in studies of this type: this wavelength regime contains a plethora of useful features, as outlined in \cite{2012ApJ...747...69C} (hereafter CvD12a). Moreover, advances in instrumentation and the development of empirical stellar spectral libraries at longer wavelengths (e.g. \citealp{2009ApJS..185..289R}) allow much more accurate measurement and modelling of these features than has previously been possible. Recent examples include \cite{2016ApJ...821...39M}, \cite{2016MNRAS.457.1468L}, \cite{2017ApJ...841...68V}, \cite{2016arXiv161200364V}, and \cite{2017MNRAS.465..192Z}, as well as the first paper in this series, \cite{2017MNRAS.468.1594A} -- hereafter cited as Paper I (details below).
	
Particular attention has been paid in these studies to spatial variations in the form of the IMF. ETGs are thought to assemble most of their stellar mass rapidly in a (swiftly quenched) starburst phase at z$\sim$2--3, before evolving passively through dry minor mergers. Observations indicate that during this latter phase ETGs greatly increase in size (e.g. \citealp{2009ApJ...697.1290B}, \citealp{2009MNRAS.398..898H}, \citealp{2010MNRAS.401.1099H}), a picture supported by simulations (e.g. \citealp{2010ApJ...725.2312O}, \citealp{2012ApJ...744...63O}, \citealp{2013MNRAS.429.2924H}). Thus, the stellar populations in the outskirts of ETGs have a different physical origin than those in the core (where, due to signal-to-noise constraints, most existing observations have been made) and may, therefore, also be formed according to a different IMF. To date, however, the observational picture remains ambiguous, with \cite{2015MNRAS.447.1033M}, \cite{2016MNRAS.457.1468L}, and \cite{2016arXiv161109859V} finding evidence for radial variations in the IMF in a number of ETGs, while in the works of \cite{2016arXiv161200364V}, \cite{2017MNRAS.465..192Z}, and in Paper I no clear evidence is found.
	
The KMOS Infrared Nearby Early-Type Survey (KINETyS) is an investigation of the stellar population properties (principally ages, chemical abundances, and underlying IMF) of nearby ETGs, with a particular focus on radial variations of these properties. In the first KINETyS paper (i.e. Paper I) we presented VLT-KMOS spectra covering the 0.78--1.34$\upmu$m range for a sample of eight nearby ETGs. Through measurements of a variety of spectral indices, in concert with archival optical data for the targets from the ATLAS$^{\rm 3D}$ survey \citep{2015MNRAS.448.3484M}, we inferred radial stellar population gradients. By stacking spectra, we were able to track the average gradients for the sample out to the effective radius. We found some strong chemical abundance gradients (e.g. [$\alpha$/H]\,=\,--0.20$\pm$0.01, [Na/H]\,=\,--0.53$\pm$0.07) but did not find evidence for radial IMF variations (variations in the fractional contribution of dwarf stars to the light were limited to a few percent at most), and were unable to constrain the abundances of some elements and stellar population age.
	
In this paper we extend the work presented in Paper I by measuring a set of gravity-sensitive spectroscopic features in the K-band. In principle this can further mitigate the degeneracies inherent in stellar population modelling, whilst also facilitating a comparison of models and data in this interesting yet still less-well-explored wavelength regime. While studies of the IMF rooted in infrared spectroscopy are becoming more common, there remains a great deal of less-explored territory at longer wavelengths ($>1\upmu$m). Spectroscopic models indicate that this regime is no-less promising than the more commonly studied bands (e.g. see CvD12a), containing a multitude of strong absorption features with dependencies on a variety of stellar population properties. 

In this work we deploy a more sophisticated approach to stellar population inference. We use an updated set of stellar population models, which are based on the updated IRTF library of \citealp{2017ApJS..230...23V} and are described in full in \citealp{2018ApJ...854..139C}. These models use a more comprehensive set of stellar population parameters than the models of Paper I. Hereafter we refer to these as the CvD16 models, and to those used in Paper I as the CvD12 models.
	
The paper is organised as follows: in Section 2 we present new K-band data from the KINETyS sample. In Section 3 we describe our procedures for measuring the strengths of various spectroscopic features. In Section 4 we present measurements of empirical spectroscopic gradients for K-band features in the KINETyS sample and compare these with predictions from the models of Paper I. In Section 5 we give details of our updated methods for stellar population inference and in Section 6 we present results thereby derived for the KINETyS sample. We discuss all our results in Section 7 and give our conclusions in Section 8.
	
\section{Data}

\subsection{Observations:}

Seven of the eight nearby ETGs introduced in Paper I were re-observed with KMOS \citep{2013Msngr.151...21S} in the K-band (1.93--2.50\umicron) using the same `sparse mosaic' observing strategy as in that work (see Fig. 1, Paper I). The data were obtained between 20$^{\rm th}$ February and 8$^{\rm th}$ of August 2016 (run ID: 097.B-0882(A), PI: Alton). Sample details are given in Table \ref{table:kmos_data}. In the K-band, KMOS has resolving power R\,=\,4200. For most targets we took 6$\times$480s exposures on each pointing, resulting in a total on-source integration time of 2880s (for NGC\,1407 and NGC\,4621 it was only possible to take half as many exposures, for a total of 1440s on-source). The signal-to-noise ratio ranges from $>$\,200 per pixel for the innermost region of the galaxies, falling to $\sim$\,10 at the effective radius (so those data are only useful when combined across the sample).
		
\begin{table*}
	\centering
	\caption{List of sample galaxies with observation details and key properties listed. Effective radii and total (J-band) magnitudes within the effective radius were extracted from 2MASS J-band images and used to calculate the mean surface brightness. Recession velocities were derived using pPXF and $\sigma(\mathrm{R_{eff}}/8)$ values and fast/slow rotator status were taken from the ATLAS 3D survey (except for NGC\,1407 for which a value was derived from our pPXF fits). Relative M/L taken from CvD12b.}
	\label{table:kmos_data}
	\begin{tabular}{cccccrccccc}
		\hline
		   Name     &   Seeing   & observation time & R$_{\mathrm{eff}}$ &          cz          & $\sigma(\mathrm{R_{eff}}/8)$ &   surface brightness    &  relative M/L   & ellipticity &    Notes     &  \\
		            &  arcsec   &        sec        &      arcsec       & $\mathrm{kms^{-1}}$ &         $\mathrm{kms^{-1}}$ & mag$_J\,$arcsec$^{-2}$ & (Milky Way = 1) &             &              &  \\ \hline
		NGC$\,$1407 & 0.68--0.75 &       1440       &        36.2        &         1950         &                          301 &          17.4           &       ---       &    0.00     & Slow rotator &  \\
		NGC$\,$3377 & 0.88--1.16 &       2880       &        23.3        &         690          &                          146 &          17.0           &      1.16       &    0.40     & Fast rotator &  \\
		NGC$\,$3379 & 0.95--1.24 &       2880       &        28.5        &         900          &                          213 &          16.5           &      1.60       &    0.00     & Fast rotator &  \\
		NGC$\,$4486 & 0.99--1.34 &       2880       &        44.5        &         1290         &                          314 &          16.9           &      1.90       &    0.00     & Slow rotator &  \\
		NGC$\,$4552 & 0.89--1.07 &       2880       &        24.1        &         390          &                          262 &          16.6           &      2.04       &    0.00     & Slow rotator &  \\
		NGC$\,$4621 & 1.02--1.22 &       1440       &        27.7        &         480          &                          224 &          16.8           &      1.96       &    0.33     & Fast rotator &  \\
		NGC$\,$5813 & 1.13--1.57 &       2880       &        33.7        &         1920         &                          226 &          17.9           &      1.37       &    0.25     & Slow rotator &  \\ \hline
	\end{tabular}
\end{table*} 

\subsection{Data reduction:}

The KMOS data were reduced mostly as in Paper I, using a slightly modified version of the ESO standard pipeline. However, we modified the atmospheric absorption correction procedure with the \textsc{MolecFit} (v1.0.2) tool. To achieve optimal performance we separately fit the atmosphere in two wavelength ranges, 19587--21075\AA\, and 21075--23964\AA, modelling H$_2$O, CH$_4$, and CO$_2$ absorption. The first of these ranges contains strong, broad absorption by CO$_2$ which makes it difficult to fit a polynomial continuum through this range. Splitting the fit into two (largely independent) segments prevents this issue from affecting the redder part of the spectrum where most of the stellar features of interest are. 

We improve our methods from Paper I by characterising the variability of the atmospheric correction in each observing run, i.e. over six exposures (at each wavelength we estimate the 68\% scatter between the set of six applied corrections). This likely represents an overestimate of the error (since, in reality, we expect that the atmospheric absorption profile will vary between observations, rather than differences being entirely due to observational uncertainties/scatter). Nevertheless we propagate this additional uncertainty to later reduction steps.

As before, spectra from the same radial extraction region were, for each galaxy, combined to create a single 1D spectrum for each physical region of the galaxy probed. This was accomplished by first dividing out continuum variations, then median-stacking the spectra.

\subsection{Velocity dispersions and index measurements}

In order to correct our measurements to a common velocity dispersion we made use of the redshifts and velocity dispersions derived in Paper I. We note that for spectra with such high intrinsic velocity dispersion, the difference in instrumental broadening between bands has a negligible effect, increasing the total velocity dispersion by $<$10\,kms$^{-1}$ and hardly affecting the correction of the index strengths.

Table \ref{table:f_defs} describes all spectral features considered in this work. This set of indices includes three optical features, for which we use archival ATLAS$^{\rm 3D}$ measurements presented in \cite{2015MNRAS.448.3484M} -- these are essential for providing good constraints on [Fe/H] and strengthen constraints on other properties. In addition, we make use of the index measurements presented in Paper I. Finally, for each spectrum we measured the equivalent widths of a set of K-band spectral indices. 

For each feature we used the flux uncertainties to create a set of Monte-Carlo realisations of our spectra from which the uncertainties of these index measurements could be inferred. The measured velocity dispersions were used to correct all index measurements to 230\,kms$^{-1}$; the uncertainty on these corrections was also propagated. All measured equivalent widths and uncertainties are given in Appendix A, Tables \ref{table:t1407} to \ref{table:t5813}.

\begin{table*}
	\centering
	\caption[absorption feature index definitions]{List of absorption index names and definitions (vacuum wavelength definitions, given in \AA); originally from CvD12a.}
	\label{table:f_defs}
	\begin{tabular}{l|ccc|l}
		\hline
		&      Blue      &    Feature     &      Red       &  \\
		Index Name                &   continuum    &   Definition   &   continuum    & Notes  \\ \hline
		H$\beta$                  & 4827.9--4847.9 & 4847.9--4876.6 & 4876.6--4891.6 & ATLAS$^{\rm 3D}$ data        \\
		Fe 5015\AA                & 4946.5--4977.8 & 4977.8--5054.0 & 5054.0--5065.3 &  \\
		Mg\textit{b}              & 5142.6--5161.4 & 5160.1--5192.6 & 5191.4--5206.4 &  \\ \hline
		Na\,I (0.82\,$\upmu$m)    & 8170.0--8177.0 & 8177.0--8205.0 & 8205.0--8215.0 & Paper I data \\
		Ca\,II (0.86$\,\upmu$m a) & 8484.0--8513.0 & 8474.0--8484.0 & 8563.0--8577.0 &  \\
		Ca\,II (0.86$\,\upmu$m b) & 8522.0--8562.0 & 8474.0--8484.0 & 8563.0--8577.0 &  \\
		Ca\,II (0.86$\,\upmu$m c) & 8642.0--8682.0 & 8619.0--8642.0 & 8700.0--8725.0 &  \\
		Mg\,I (0.88$\,\upmu$m)    & 8801.9--8816.9 & 8777.4--8789.4 & 8847.4--8857.4 &  \\
		FeH (0.99$\,\upmu$m)      & 9905.0--9935.0 & 9855.0--9880.0 & 9940.0--9970.0 &  \\
		Ca\,I (1.03$\,\upmu$m)    &  10337--10360  &  10300--10320  &  10365--10390  &  \\
		Na\,I (1.14$\,\upmu$m)    &  11372--11415  &  11340--11370  &  11417--11447  &  \\
		K\,I (1.17$\,\upmu$m a)   &  11680--11705  &  11667--11680  &  11710--11750  &  \\
		K\,I (1.17$\,\upmu$m b)   &  11765--11793  &  11710--11750  &  11793--11810  &  \\
		K\,I (1.25$\,\upmu$m)     &  12505--12545  &  12460--12495  &  12555--12590  &  \\
		Al\,I (1.31$\,\upmu$m)    &  13115--13165  &  13090--13113  &  13165--13175  &  \\ \hline
		Ca\,I (1.98$\,\upmu$m a)  &  19740--19765  &  19770--19795  &  19800--19840  & new KMOS data  \\
		Ca\,I (1.98$\,\upmu$m b)  &  19800--19840  &  19845--19880  &  19885--19895  &  \\
		Na\,I (2.21$\,\upmu$m)    &  22035--22045  &  22047--22105  &  22107--22120  &  \\
		Ca\,I (2.26$\,\upmu$m)    &  22500--22575  &  22580--22700  &  22705--22780  &  \\
		CO (2.30$\,\upmu$m a)     &  22860--22910  &  22932--22982  &  23020--23070  &  \\
		CO (2.30$\,\upmu$m b)     &  23150--23200  &  23220--23270  &  23300--23350  &  \\ \hline
	\end{tabular}
\end{table*}

\subsection{Stacked spectra}

As in Paper I we chose to make use of stacked spectra in our analysis, creating these by the same method as before (described below). This has a number of advantages, the most important being that stacking spectra with slightly different redshifts in the source rest-frame suppresses any systematics introduced in the observed rest-frame (e.g. due to poor subtraction of sky emission lines or sub-par telluric corrections). Additionally, stacked spectra can be used to constrain the average behaviour of the sample (and, in turn, this can be used to identify unusual objects that deviate from this average).

For the centrally extracted KMOS spectra, the stacking procedure was as follows: we shifted the input spectra into the rest-frame and binned them onto a common wavelength grid. We characterised the relative continuum variation by dividing each spectra by the mean spectrum and fitting a 6$^{\rm th}$ order polynomial to the ratio of the two. This relative continuum variation was then divided out and the median flux at each wavelength was evaluated. Errors were created by bootstrap resampling of the input spectra (since the scatter between galaxies is larger than the statistical uncertainty on any given spectrum).

\begin{figure*}
	\centering
	\includegraphics[width=0.99\textwidth]{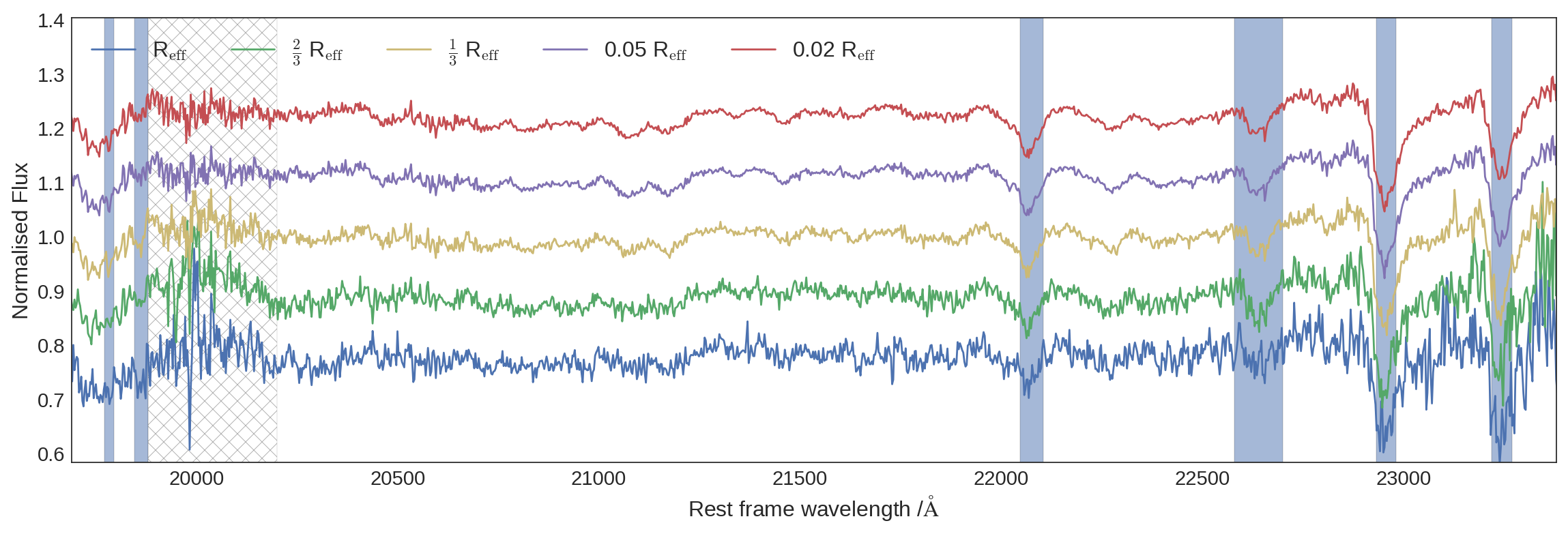}
	\caption[KMOS K-band stacked spectra]{K-band spectra from KMOS, median-stacked at fixed fraction of the effective radius. Even in the outlying regions of the galaxy the signal of the strong absorption features are recovered via this method. The hatched region shows a region of particularly strong telluric absorption (atmospheric transmission\,$\lesssim$\,50\%). The features listed in Table \ref{table:f_defs} are shown by the light blue bands.}
	\label{fig:KMOS_stacks}
\end{figure*}    

We divided the other KMOS spectra by the centrally-extracted spectrum prior to stacking, so that we stacked the \textit{radial variations} of the spectra. We then multiplied these `variation stacks' by the stacked spectrum already derived from the central extraction region. This procedure maximises the radial variation signal: any constant offset in the strength of spectral features between different galaxies in the stack is divided out.

The KMOS K-band spectra are shown in Fig. \ref{fig:KMOS_stacks}. We measured equivalent widths and statistical uncertainties of available spectroscopic features just as for the spectra of individual galaxies. These measurements are also provided in Appendix A, in Table \ref{table:kmos_k_stacks}.
	
\section{Spectroscopic radial gradients}

\subsection{Interpretation of absorption line strengths}

As outlined in the Introduction, absorption features in stellar population spectra encode information about the properties of the stars as well as kinematic information (recession velocities, velocity broadening). The strength of those spectral features in the light of individual stars depends on the chemical composition, age, and surface gravity of the star. Because of these dependencies it is in principle possible to constrain the chemical abundance pattern, population age, and relative contribution of dwarf and giant stars to the total light (and thus the IMF).

We present the strengths of four K-band features (definitions are given in Table \ref{table:f_defs}) in seven ETGs at five radial locations. These features are as follows:

\subsubsection*{Ca\,I 1.98\,$\mu$m}

The comparatively weak Ca\,I doublet is principally sensitive to Ca abundance and the IMF, becoming stronger as Ca abundance increases and as the IMF becomes more bottom-heavy. This feature was first highlighted in CvD12a as an IMF-sensitive index with the potential to help constrain the shape of the low-mass IMF, but to our knowledge has not been measured before. In Appendix B we show that (given other constraints on f$_{\rm dwarf}$) Ca\,I 1.98\umicron\, can indeed provide information about the IMF shape. This feature lies next to a band of strong telluric absorption and so in some cases is difficult to measure with precision. 

\subsubsection*{Na\,I 2.21\,$\mu$m}

This strong feature is, in principle, extremely useful for constraining [Na/Fe], since its sensitivity to the IMF is comparatively limited, in contrast to the Na\,I lines at 0.82$\upmu$m and 1.14$\upmu$m. Previous work (\citealp{2015MNRAS.454L..71S} and see also Paper I) has indicated that in the CvD12 models, the strengths of these lines are in tension with each other, given other constraints. The measurement of other strong Na features may help to resolve this tension: although the optical Na\,I 0.59$\upmu$m line would be more readily measurable and even less sensitive to the IMF, the Na\,I 2.21$\upmu$m line can nonetheless provide a valuable constraint. Like the 0.82$\upmu$m and 1.14$\upmu$m lines, Na\,I 2.21$\upmu$m is unaffected by interstellar absorption from dust and gas.

This feature has been studied previously at an empirical level in non-spatially resolved spectra by \cite{2008ApJ...674..194S} and \cite{2009ApJ...705L.199M}. Although at the time accurate SSP models were not available in the K-band, Na\,I 2.21\umicron\, was noted to correlate strongly and positively with galaxy velocity dispersion and also to be very strong in comparison to measurements of Milky Way stars. This latter result was also found by \cite{2008ApJ...677..238D}, in which radial variations of this feature in the central 1\arcsec\, of compact elliptical galaxies were studied. \cite{2015A&A...582A..97M} compared these strengths with new K-band SSP models, confirming that these generally predicted lower strengths for Na\,I 2.21\umicron\, than had been measured. A more recent treatment by \cite{2017MNRAS.464.3597L} using a new version of the MILES SSP models studied this and other Na features in the optical and near infrared, finding that the strength of the feature could only be reproduced by a combination of Na superabundance \textit{and} a bottom-heavy IMF.

\subsubsection*{Ca\,I 2.26\,$\mu$m}

This strong line is not particularly sensitive to the IMF, being principally sensitive to the Ca abundance in the CvD16 models, with minor contributions from other elements. \cite{2008ApJ...674..194S} found that it did not correlate strongly with galaxy velocity dispersion, but showed significant scatter, while \cite{2009A&A...497...41C} reported a loose positive correlation. Ca\,I 2.26\umicron\, has not been well-studied using SSP modelling.

\subsubsection*{CO 2.30\,$\mu$m bandhead}

The extremely strong CO molecular absorption bands at the red end of the K-band are most prominent in giant stars -- and so decline in strength if the IMF is bottom-heavy. A competing effect comes from [C/Fe], an increase in which will, of course, strengthen these features. The CO bandhead has in the past been used to measure stellar kinematics (e.g. \citealp{2008A&A...485..425L}, \citealp{2011MNRAS.412.2017V}) as well finding use in discussions of stellar populations at an empirical level, e.g. \cite{2008ApJ...674..194S}, \cite{2009A&A...497...41C}. Both these studies found significant correlation between CO and measurements of optical features. \cite{2008ApJ...677..238D} studied radial variation in this feature's strength in the central 1\arcsec of compact ellipticals, finding evidence for a complex radial behaviour in some galaxies; the strength rises out to R\,=\,0.3\arcsec, then falls for R\,$>$\,0.3\arcsec. \cite{2009ApJ...705L.199M} discussed the CO bandhead in the context of stellar population models, arguing that it could be used to test for extended star-formation histories in ETGs by probing the contribution to the light of intermediate-age stars.

\subsection{Absorption feature gradients}

In Paper I we were able to constrain the stellar population gradients in our sample, finding evidence for strong chemical abundance gradients that seemed to be fairly consistent across our sample. We now present radial gradients in the strengths of the K-band absorption lines we have measured (the first time these gradients have been measured, in some cases) and compare these with the predictions of the best-fit model from Paper I.

\begin{figure*}
	\centering
	\includegraphics[width=0.98\textwidth]{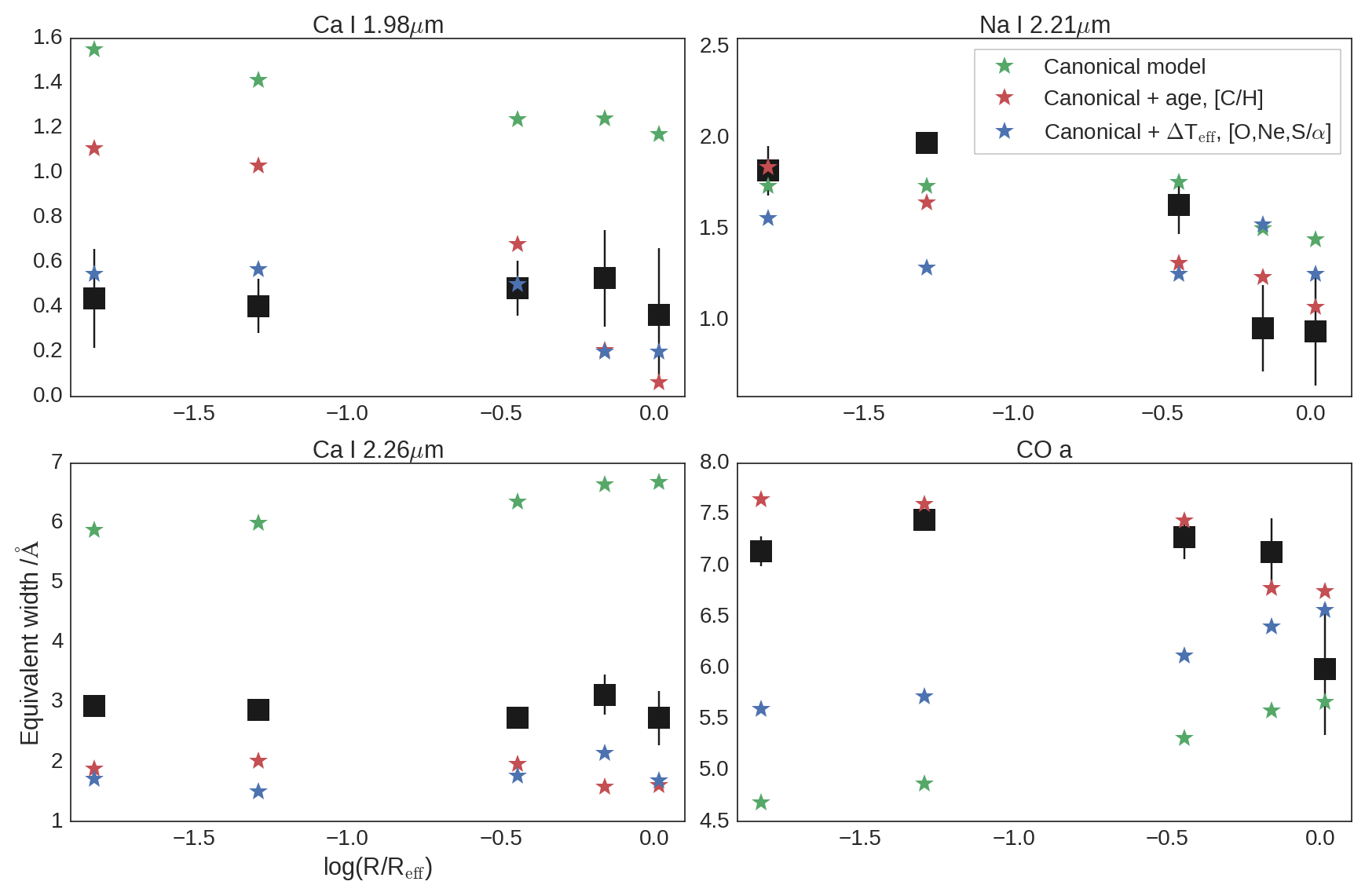}
	\caption[K-band absorption line gradients and comparison with model predictions]{
		Measurements from the stacked KMOS K-band spectra with errors (large black symbols). Also shown are the predictions of the best-fit models from Paper I, which were \textit{not} constrained using these data (i.e. these are true predictions). \\ \textit{Green:} the `canonical model' of Paper I (which accounts for [$\alpha$/H], [Fe/H], [Na/H], [Ca/$\alpha$], [K/H], as well as the IMF). \\ \textit{Red:} an extended model which additionally accounts for stellar population age and [C/H] (the inclusion of both proved necessary to fit the optical H$\beta$ line measurements taken from ATLAS$^{\rm 3D}$). \\ \textit{Blue:} an extended version of the canonical model with the fitting process marginalised over the nuisance parameters $\Delta$T$_{\rm eff}$ and [O,\,Ne,\,S/$\alpha$]. This made little difference in the fits to the IZ \& YJ band data.
	}
	\label{fig:k_stack_measurements}
\end{figure*}

In Fig. \ref{fig:k_stack_measurements} we show the predictions derived from the models used in Paper I (i.e. the CvD12 models) against the new measurements, given the best-fit parameters we found previously. These predictions indicate that the strength of the Ca features should be approximately constant with radius, which the data bear out. However, the `canonical' model of Paper I severely overpredicts the strengths of the Ca\,I features. The canonical model used the minimum set of stellar population parameters required to explain the optical, IZ-, and YJ-band data adequately. That model also underpredicts the CO bandhead $a$ line, but does not account for variations in the abundance of carbon, which the CO bandhead's strength crucially depends on. In Paper I we explored a model with a [C/H] parameter included, finding this necessary to account for the low strength of the H$\beta$ line. While the value of this parameter was not very well constrained, this model is much less discrepant with the CO bandhead's strength, suggesting good prospects for constraining the carbon abundance in the updated model framework. We also explored a model that incorporated two additional nuisance parameters, $\Delta$T$_{\rm eff}$ and [O,\,Ne,\,S/$\alpha$]; this model produced much better estimates for the Ca\,I lines. Finally, the Na\,I 2.21\umicron\, line exhibits a steeper gradient than is predicted in any of the models used in Paper I (despite the steep [Na/H] gradient we inferred). 

The model sensitivities of these features and good S/N of some our measurements indicate that these data will add significant additional constraining power to our stellar population analysis. In Table \ref{table:line_strength_gradients} we give the best-fit empirical spectroscopic gradients derived for these features (via the same procedure as in Paper I), with associated statistical uncertainties.

\begin{table}
	\centering
	\caption[K-band best-fit feature gradients]{Best-fit radial gradients (change in equivalent width per decade in radius) derived from the KMOS stacks.}
	\begin{tabular}{cc}
		\hline
		Feature                & best-fit gradient (\AA) per dex in log(R/R$_{\rm eff}$)\\ \hline
		Ca\,I 1.98\umicron           &      \gp{}0.05$\pm$0.02      \\
		Na\,I 2.21\umicron           &     --0.53$\pm$0.08      \\
		Ca\,I 2.26\umicron           &     --0.08$\pm$0.09      \\
		CO 2.30\umicron\, $a$           &     --0.11$\pm$0.13      \\ \hline
		\label{table:line_strength_gradients} &
	\end{tabular}
\end{table}

The strengths of the Na\,I 2.21\umicron\, feature and the CO $a$ line appear to flatten/turn over in the innermost region, where two spectra were extracted from the central IFU for each galaxy. It is important to understand whether this issue results from real variation of the stellar populations or some unaccounted-for observational effect. An instrumental issue seems unlikely, as the two data points are extracted from the same IFU. Furthermore, the effect is not seen in the Ca\,I 2.26\umicron\, feature. The effects of observational seeing can in principle reduce the difference between the two measurements by spreading light from one region into another; we expect this effect to be limited since the seeing for these observations was typically $<$1.4\arcsec\, (the central extraction region has this diameter). Nor is the seeing for these observations notably worse (on average) than for the IZ and YJ observations we made.

\section{Stellar population analysis}

\subsection{Updated model implementation}

For this work, as in Paper I, model parameter estimation is accomplished using the \textsc{emcee} routine \citep{2013PASP..125..306F}. This is a python implementation of an affine-invariant Markov-Chain Monte Carlo (MCMC) ensemble sampler (see \citealp{Goodman_and_Weare}) used to characterise the posterior probability distribution for a chosen model, given data and statistical uncertainties. 

\begin{figure*}
	\centering
	\includegraphics[width=0.99\textwidth]{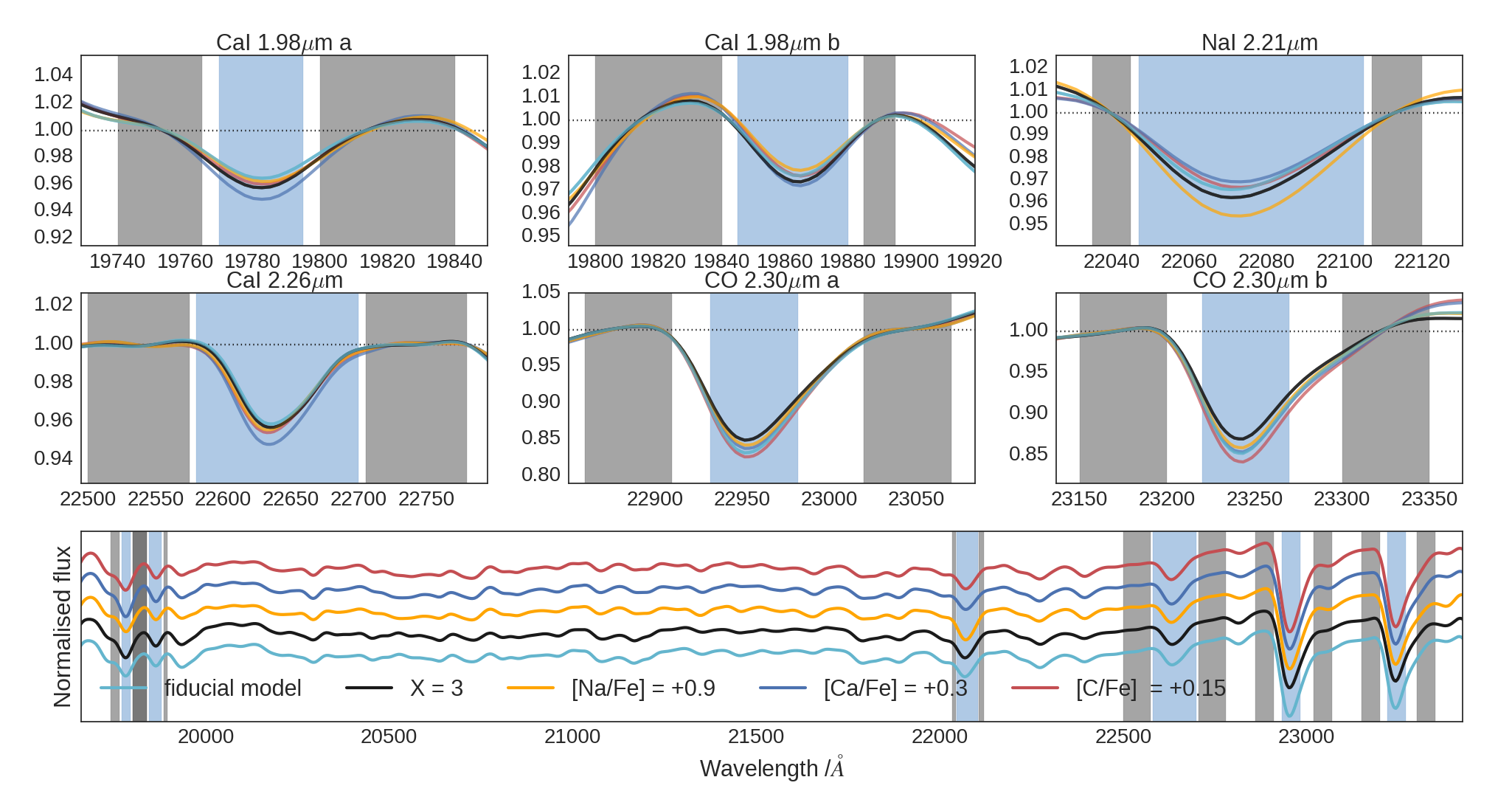}
	\caption[K-band spectral models]{A subset of the CvD16 models used in this work broadened to a velocity dispersion of 230\,kms$^{-1}$, indicating the effect of particularly important parameters. The fiducial model is for a 13.7\,Gyr old population formed via a Kroupa IMF, with a solar abundance pattern. \\ \textit{Lower panel:} full models for the K-band with measured feature bands indicated in blue and associated pseudo-continuum bands in grey. \\ \textit{Upper panels:} close-up view of each measured feature, with the models normalised with respect to the feature pseudo-continua.}
	\label{fig:k_models}
\end{figure*}

For this purpose we here make use of an updated version of the CvD12 models used in Paper I. The CvD16 models are constructed from an expanded library of stellar spectral templates incorporating stars with a variety of metallicities \citep{2017arXiv170508906V}, spanning the range [Z/H] = --1.5 to [Z/H] = +0.2. In Fig. \ref{fig:k_models} we show a few of these models corresponding to the K-band spectra of different stellar populations with different chemical abundance patterns and IMFs.

The CvD16 models allow more freedom in the IMF parametrization. In Paper I, IMF variations were parametrized with a single parameter, f$_{\rm dwarf}$, while the underlying shape of the IMF was assumed to be a single power-law below 1\,M$_\odot$. \fdwarf was defined as the fractional (J-band) luminosity contributed by stars of mass less than 0.5\,M\,$_\odot$.

In Paper I we showed that for the spectral features used in our earlier analysis, index strength variations are approximately linear with respect to the f$_{\rm dwarf}$ parameter, with little dependence on the IMF shape. However, the models indicate that the effect of IMF shape is \textit{not} negligible for the K-band indices we have measured here. This is because the strengths of different features trace the contribution to the light from stars in different mass ranges. Full details are given in Appendix B.

Because of this dependence on the IMF shape, in this work we use models with additional freedom in the IMF parametrization. Therefore, in this work the IMF is modelled as a broken power law, comprising two sections (0.08\,M$_{\odot}$\,--\,0.5\,M$_{\odot}$ and 0.5\,M$_{\odot}$\,--\,1.0\,M$_{\odot}$) with independent power-law slopes X1 and X2. In this formalism, the Kroupa IMF (i.e. the Milky Way IMF) can be specified by X1\,=\,1.3, X2\,=\,2.3.

The CvD16 models also make use of a different approach to modelling the chemical abundance pattern of the stellar populations to the CvD12 models. The overall metallicity of the population is now represented using the relative logarithmic abundance [Z/H] (where [Z/H]\,=\,0 corresponds to the solar abundance pattern), while individual variations in element abundances are computed with respect to the overall metallicity (i.e. they are parameterised as [X/Z]). In addition, the response of the index strengths to variations in the abundance of particular elements varies with [Z/H] in the updated models, meaning the responses are now coupled.

The full set of parameters used in this work for population inference is given in Table \ref{table:prior_table2}, along with the cut-off limits of the prior distribution.

\begin{table}
	\centering
	\caption[MCMC model parameters and prior probability distributions]{List of model parameters used in the MCMC inference. X1 and X2 are the IMF power-law slopes over the stellar mass intervals 0.08--0.5 and 0.5--1.0M$_{\odot}$. Z is the total metallicity of the population and the [X/Z] parameters are the relative variations of individual elements with respect to the overall metallicity.}
	\begin{tabular}{ccc}
		\hline
		Model parameter & Minimum value & Maximum value \\
		\hline
		log(t$_{\rm age}$/Gyr) & log(7.0) & log(14.0) \\
		X1    & 1.0   & 3.5 \\
		X2    & 1.0   & 3.5 \\
		$\rm [Z/H]$   & --1.0 & 0.4 \\
		$\rm [Fe/Z]$ & --0.5 & 0.3 \\
		$\rm [Mg/Z]$ & --0.4 & 0.6 \\
		$\rm [Ca/Z]$ & --0.4 & 0.4 \\
		$\rm [Ti/Z]$ & --0.4 & 0.6 \\
		$\rm [O,\,Ne,\,S/Z]$ & --0.3 & 0.3 \\
		$\rm [Na/Z]$ & --0.6 & 0.9 \\
		$\rm [K/Z]$  & --0.3 & 0.3 \\
		$\rm [C/Z]$  & --0.3 & 0.3 \\
		\hline
		\label{table:prior_table2}
	\end{tabular}
\end{table}

For the purpose of presenting our results, some of the model parameters are remapped to alternative quantities for clarity and ease of comparison with other works. Stellar population age, rather than log(t$_{\rm age}$) is given. The segmented IMF slopes are converted to f$_{\rm dwarf}$ and also to the relative V-band M/L ratio. The latter describes the ratio of M/L for two populations, one formed via the chosen IMF and the other according to a Kroupa IMF (but, to aid comparison, without adjusting for metallicity or age; we assume [Z/H] = 0 and t$_{\rm age}$=13.5\,Gyr). To describe the abundances, we compute [Fe/H] = [Fe/Z] + [Z/H] and, for each of the other elements, [X/Fe] = [X/Z] -- [Fe/Z]. These quantities are further adjusted to account for measured deviations in [X/Fe] from the solar ratio in the library stars of different metallicities using the same recipe as was employed in \cite{2017ApJ...841...68V}.

\subsection{Comparison with previous model implementation}

To allow a meaningful comparison, we first use these models to fit the optical, IZ-, and YJ-band stacked data presented in Paper I in order to (a) reproduce the results already given and (b) highlight any differences due to the model changes prior to including the information from our new K-band measurements, i.e. isolate the effect of updating the model framework.

To achieve this, we set up the models with the constraint X1\,=\,X2 (i.e. we limited ourselves to the case of a single power-law description of the IMF). We assumed an old age (t\,=\,13.5\,Gyr) and included measurements of the optical features Fe\,5015 and Mg\textit{b} in the fit. We fit for metallicity, the IMF slope, and abundance variations in Fe, Mg, Na, Ca, and K. The results of a set of MCMC runs with 100 walkers and 900 steps (after 100 burn-in steps) are given in Table \ref{table:basic_results}.

\begin{table*}
	\centering
	\caption[comparison of updated models with preceding version]{Estimated parameter values for the stacked spectra first presented in Paper I, acquired using the CvD16 models. We set X1\,=\,X2\,=\,X and fit for [Z/H] as well as the abundances of various elements of interest. R1 and R2 correspond to the two extraction regions in the central IFU ($<0.7\arcsec$, $>0.7\arcsec$) while R3, R4, and R5 correspond to the rings of IFUs arranged at $\sim1/3\,$R$_{\rm eff}$, $\sim2/3\,$R$_{\rm eff}$, and $\sim$\,R$_{\rm eff}$.}
	\begin{tabular}{lcccccc}
		\hline
		&            R1             &            R2             &            R3             &            R4             &            R5             &      gradient      \\ \hline
		X                    & \gp{}1.07\err{0.50}{0.05} & \gp{}1.07\err{0.51}{0.05} & \gp{}2.16\err{0.55}{0.74} & \gp{}1.22\err{1.07}{0.14} & \gp{}2.71\err{0.53}{1.10} & \gp{}0.49$\pm$0.36 \\
		$\rm [Z/H]$          & \gp{}0.24\err{0.04}{0.05} & \gp{}0.14\err{0.03}{0.04} &  --0.03\err{0.06}{0.11}   &  --0.07\err{0.09}{0.13}   &  --0.15\err{0.13}{0.15}   &  --0.22$\pm$0.05   \\
		$\rm [Fe/H]$         &  --0.07\err{0.07}{0.06}   &  --0.19\err{0.05}{0.07}   &  --0.26\err{0.08}{0.12}   &  --0.31\err{0.12}{0.13}   &  --0.25\err{0.14}{0.18}   &  --0.13$\pm$0.04   \\
		$\rm [Mg/Fe]$        & \gp{}0.41\err{0.06}{0.05} & \gp{}0.44\err{0.05}{0.05} & \gp{}0.40\err{0.09}{0.07} & \gp{}0.43\err{0.10}{0.12} & \gp{}0.28\err{0.15}{0.14} &  --0.03$\pm$0.03   \\
		$\rm [Na/Fe]$        & \gp{}0.84\err{0.09}{0.09} & \gp{}0.73\err{0.06}{0.07} & \gp{}0.43\err{0.15}{0.27} & \gp{}0.35\err{0.27}{0.38} &  --0.12\err{0.47}{0.31}   &  --0.38$\pm$0.13   \\
		$\rm [Ca/Fe]$        & \gp{}0.17\err{0.08}{0.08} & \gp{}0.16\err{0.07}{0.06} & \gp{}0.11\err{0.10}{0.10} & \gp{}0.04\err{0.14}{0.14} &  --0.01\err{0.15}{0.21}   &  --0.08$\pm$0.04   \\
		$\rm [K/Fe]$         & \gp{}0.16\err{0.12}{0.14} & \gp{}0.11\err{0.11}{0.10} & \gp{}0.06\err{0.23}{0.18} & \gp{}0.16\err{0.22}{0.23} & \gp{}0.19\err{0.27}{0.30} &  --0.00$\pm$0.08   \\ \hline
		f$_{\mathrm{dwarf}}$ & \gp{}2.2\err{1.5}{0.2}\%  & \gp{}2.2\err{1.6}{0.1}\%  & \gp{}2.9\err{7.0}{0.6}\%  & \gp{}2.8\err{6.0}{0.5}\%  & \gp{}3.2\err{13.0}{0.9}\% & \gp{}0.3$\pm$2.7\% \\
		M/L                  & \gp{}0.70\err{0.18}{0.02} & \gp{}0.69\err{0.18}{0.02} & \gp{}0.79\err{0.96}{0.08} & \gp{}0.78\err{0.82}{0.08} & \gp{}0.86\err{2.04}{0.13} & \gp{}0.04$\pm$0.35 \\ \hline
	\end{tabular}
	\label{table:basic_results}
\end{table*}	

\begin{table*}
	\centering
	\caption[reparameterised results from Chapter 4]{The results presented in Paper I with the parameters reorganised to match those presented in Table \ref{table:basic_results}.}
	\begin{tabular}{lcccccc}
		\hline
		&            R1             &            R2             &            R3             &            R4             &            R5             &    gradient     \\ \hline
		$\rm [Fe/H]$      & \gp{}0.06\err{0.03}{0.03} &  --0.06\err{0.03}{0.03}   &  --0.21\err{0.05}{0.05}   &  --0.21\err{0.08}{0.08}   &  --0.21\err{0.10}{0.10}   & --0.17$\pm$0.02 \\
		$\rm [\alpha/Fe]$ & \gp{}0.34\err{0.04}{0.04} & \gp{}0.34\err{0.04}{0.04} & \gp{}0.33\err{0.06}{0.06} & \gp{}0.31\err{0.09}{0.09} & \gp{}0.21\err{0.12}{0.12} & --0.03$\pm$0.02 \\
		$\rm [Na/Fe]$     & \gp{}0.62\err{0.12}{0.12} & \gp{}0.53\err{0.08}{0.08} & \gp{}0.21\err{0.16}{0.19} & \gp{}0.21\err{0.25}{0.35} &  --0.15\err{0.45}{0.50}   & --0.31$\pm$0.07 \\
		$\rm [Ca/Fe]$     &  --0.02\err{0.06}{0.06}   & \gp{}0.00\err{0.06}{0.06} & \gp{}0.02\err{0.08}{0.08} &  --0.10\err{0.13}{0.13}   &  --0.22\err{0.18}{0.08}   & --0.03$\pm$0.02 \\
		\fdwarf           & \gp{}9.4\err{2.1}{2.1}\%  & \gp{}11.1\err{1.5}{1.5}\% & \gp{}10.5\err{2.1}{2.2}\% & \gp{}6.7\err{3.9}{1.6}\%  & \gp{}8.0\err{5.6}{2.5}\%  & --0.6$\pm$0.8\% \\ \hline
	\end{tabular}
	\label{table:reorgresults}
\end{table*}	

These results are broadly consistent with those presented in Paper I, Table 9. For clarity we reorganise those results so that they are reported in terms of the same parameters as in Table \ref{table:basic_results} and can thus be directly compared (see Table \ref{table:reorgresults}). We detect a very similar gradient in [Fe/H] and find [Mg/Fe] strongly enhanced and constant with radius. We also now explicitly calculate the [Z/H] gradient: previously it was implicitly assumed that [Z/H] = 0 at all radii. However, applying the empirical relationship [Z/H] = [Fe/H] + 0.94\,[$\alpha$/Fe] set out in \cite{2003MNRAS.339..897T} would indicate a gradient of --0.20$\pm$0.03, consistent with the new result. Meanwhile [Ca/Fe] is constant with radius and it's clear that Ca is much less over-abundant than Mg; the new model does however prefer a modest enhancement in the core, which was not the case previously. For [Na/Fe] we infer a stronger enhancement than in Paper I, although the two results are not strongly in tension and we find a consistent gradient. Finally, the IMF we infer is more Milky Way-like, rather than modestly bottom-heavy as found previously. In summary, using the updated models we find qualitatively similar results to those which were presented in Paper I.

These results are more readily compared with the literature than in the form presented in Paper I. It is instructive to make a comparison with the results of \cite{2015MNRAS.448.3484M}, which were calculated via the models of \cite{2007ApJS..171..146S} using the same optical data we used for our analysis (and assuming no IMF variations). In the central 0.125\,R$_{\rm eff}$ of the galaxies from our sample (excluding NGC\,1407, which they did not observe) the median age they find is 13.7$\pm$2.1\,Gyr, the median [Z/H] = 0.06$\pm$0.05, and the median [$\alpha$/Fe] is 0.28$\pm$0.05. The latter is a little lower than our value, which may be due to the different models used, the extra indices we fit, the use of additional parameters we include, or a combination of all of these.

In the next section we present the results of our full analysis, making use of the K-band data as well as the optical/IZ/YJ measurements, additional stellar population parameters, and the data from individual galaxies.

\section{Results}

We now use the CvD16 models in conjunction with all available data, including the KMOS measurements of the K-band indices, to infer stellar population parameters for our sample. As well as the data used in Section 4, we make use of the K-band data and include H$\beta$ in the fit: in Paper I we showed that H$\beta$'s sensitivity to various poorly constrained parameters such a [C/H] and stellar age reduced its constraining power substantially. However, with the additional sensitivity to these parameters provided by the CO bandhead our ability to constrain these parameters improves somewhat, unlocking H$\beta$'s potential as a constraint on the age of the stellar population.

While we do not expect to be able to constrain the abundance of $\alpha$ elements other than Mg and Ca, we nevertheless marginalise over the parameters [Ti/Fe] and [O,\,Ne,\,S/Fe].

\begin{table*}
	\centering
	\caption[headline results: parameter values derived from stacked spectra]{Headline results: estimated parameter values for the stacked spectra, acquired using the CvD16 models and the full set of index measurements. For the IMF we fit a two-part power law with slopes X1 and X2, and we fit for stellar population age and total metallicity, as well as the abundances of various elements of interest. R1 and R2 correspond to the two extraction regions in the central IFU ($<0.7\arcsec$, $>0.7\arcsec$) while R3, R4, and R5 correspond to the rings of IFUs arranged at $\sim1/3\,$R$_{\rm eff}$, $\sim2/3\,$R$_{\rm eff}$, and $\sim$\,R$_{\rm eff}$.}
	\begin{tabular}{lcccccc}
		\hline
		                      &            R1             &            R2             &            R3             &            R4             &            R5             &      gradient      \\ \hline
		age /Gyr              &  \gp{}11.3\err{1.4}{1.3}  &  \gp{}9.1\err{1.0}{0.7}   &  \gp{}7.8\err{2.2}{0.5}   &  \gp{}7.3\err{3.3}{0.2}   &  \gp{}8.9\err{2.5}{1.2}   &   --1.8$\pm$0.8    \\
		X1                    & \gp{}2.01\err{0.51}{0.67} & \gp{}1.10\err{0.67}{0.07} & \gp{}1.12\err{0.86}{0.08} & \gp{}1.98\err{0.66}{0.65} & \gp{}2.86\err{0.41}{1.09} & \gp{}0.26$\pm$0.39 \\
		X2                    & \gp{}1.24\err{0.90}{0.16} & \gp{}1.06\err{0.43}{0.05} & \gp{}1.49\err{0.87}{0.32} & \gp{}1.52\err{0.96}{0.35} & \gp{}1.82\err{1.02}{0.53} & \gp{}0.31$\pm$0.40 \\
		$\rm [Z/H]$           & \gp{}0.17\err{0.03}{0.04} & \gp{}0.23\err{0.02}{0.02} & \gp{}0.14\err{0.04}{0.05} & \gp{}0.02\err{0.07}{0.11} &  --0.14\err{0.13}{0.11}   &  --0.11$\pm$0.03   \\
		$\rm [Fe/H]$          &  --0.12\err{0.07}{0.11}   &  --0.08\err{0.05}{0.06}   &  --0.26\err{0.14}{0.07}   &  --0.33\err{0.15}{0.13}   &  --0.30\err{0.20}{0.18}   &  --0.16$\pm$0.05   \\
		$\rm [Mg/Fe]$         & \gp{}0.48\err{0.10}{0.05} & \gp{}0.44\err{0.05}{0.05} & \gp{}0.42\err{0.08}{0.10} & \gp{}0.51\err{0.12}{0.14} & \gp{}0.38\err{0.17}{0.17} &  --0.01$\pm$0.04   \\
		$\rm [Na/Fe]$         & \gp{}0.77\err{0.12}{0.10} & \gp{}0.70\err{0.06}{0.05} & \gp{}0.43\err{0.13}{0.16} & \gp{}0.34\err{0.21}{0.35} &  --0.11\err{0.40}{0.31}   &  --0.35$\pm$0.09   \\
		$\rm [Ca/Fe]$         & \gp{}0.20\err{0.11}{0.07} & \gp{}0.09\err{0.06}{0.05} & \gp{}0.09\err{0.11}{0.09} & \gp{}0.11\err{0.17}{0.14} & \gp{}0.08\err{0.16}{0.25} &  --0.04$\pm$0.05   \\
		$\rm [O,\,Ne,\,S/Fe]$ & \gp{}0.78\err{0.13}{0.19} & \gp{}0.64\err{0.09}{0.10} & \gp{}0.73\err{0.14}{0.15} & \gp{}0.70\err{0.19}{0.23} & \gp{}0.73\err{0.19}{0.40} & \gp{}0.02$\pm$0.06 \\
		$\rm [Ti/Fe]$         & \gp{}0.53\err{0.18}{0.16} & \gp{}0.22\err{0.12}{0.10} & \gp{}0.18\err{0.19}{0.25} & \gp{}0.57\err{0.23}{0.33} & \gp{}0.39\err{0.31}{0.46} &  --0.05$\pm$0.11   \\
		$\rm [C/Fe]$          & \gp{}0.11\err{0.10}{0.14} & \gp{}0.07\err{0.05}{0.08} & \gp{}0.15\err{0.12}{0.13} & \gp{}0.22\err{0.16}{0.18} & \gp{}0.09\err{0.24}{0.20} & \gp{}0.06$\pm$0.04 \\
		$\rm [K/Fe]$          & \gp{}0.19\err{0.17}{0.13} & \gp{}0.07\err{0.11}{0.06} & \gp{}0.19\err{0.15}{0.16} & \gp{}0.25\err{0.21}{0.20} & \gp{}0.17\err{0.29}{0.29} & \gp{}0.06$\pm$0.05 \\ \hline
		f$_{\mathrm{dwarf}}$  & \gp{}3.9\err{2.8}{0.9}\%  & \gp{}2.7\err{1.3}{0.4}\%  & \gp{}4.9\err{2.0}{1.5}\%  & \gp{}5.7\err{3.0}{2.0}\%  & \gp{}6.5\err{5.7}{2.2}\%  & \gp{}2.0$\pm$1.7\% \\
		M/L                   & \gp{}0.91\err{0.37}{0.12} & \gp{}0.75\err{0.16}{0.05} & \gp{}0.95\err{0.28}{0.14} & \gp{}1.04\err{0.45}{0.21} & \gp{}1.32\err{0.85}{0.35} & \gp{}0.19$\pm$0.13 \\ \hline
	\end{tabular}
	\label{table:K_results}
\end{table*}

\begin{figure}
	\centering
	\includegraphics[width=0.4\textwidth]{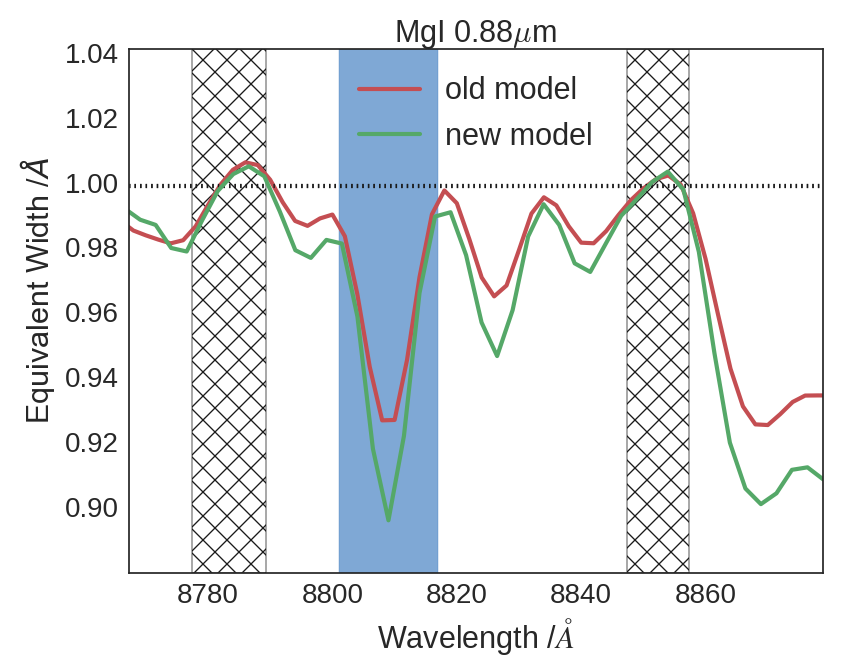}
	\caption[increased strength of Mg\,I 0.88\umicron\, in updated models]{We show a fiducial case for the old (CvD12) and new (CvD16) SSP model grids, with a Kroupa IMF, solar abundance pattern, and a stellar population age of 13.5\,Gyr. In the updated model grids (which we remind the reader are constructed from an expanded empirical spectral library) the Mg\,I 0.88\umicron\, feature (the region around which we show here alongside feature and pseudo-continuum definitions) is much stronger.}
	\label{fig:mod_comparison}
\end{figure}

To obtain these results we made use of all available feature measurements, with two exceptions. First, we found that in the updated model framework the predicted MgI 0.88\,$\upmu$m line is much stronger, given the same stellar population parameter values. This puts the new model prediction in tension with the data (we show the change to the models in Fig. \ref{fig:mod_comparison}). Secondly, we were concerned that the CO 2.30\,$\upmu$m $b$ line might be subject to much greater contamination (from a combination of increasing thermal background, stronger telluric absorption, and sky emission lines) than the CO 2.30\umicron\, $a$ line, although in the models the strengths of the two are strongly coupled. We therefore used only the $a$ line in the fit. For this strong feature the formal statistical uncertainty of the equivalent width is small, so we lose very little constraining power by doing this.

In Table \ref{table:K_results} we give the best-fit stellar population parameters inferred from the stacked spectra and provide best-fit parameter gradients with associated statistical uncertainty. All MCMC runs used 100 walkers and 900 steps after a 100 step burn-in.

\subsection{Comparison with results from Paper I}

\begin{landscape}
	\begin{figure}
		\centering
		\vskip 20mm
		\includegraphics[scale=0.39]{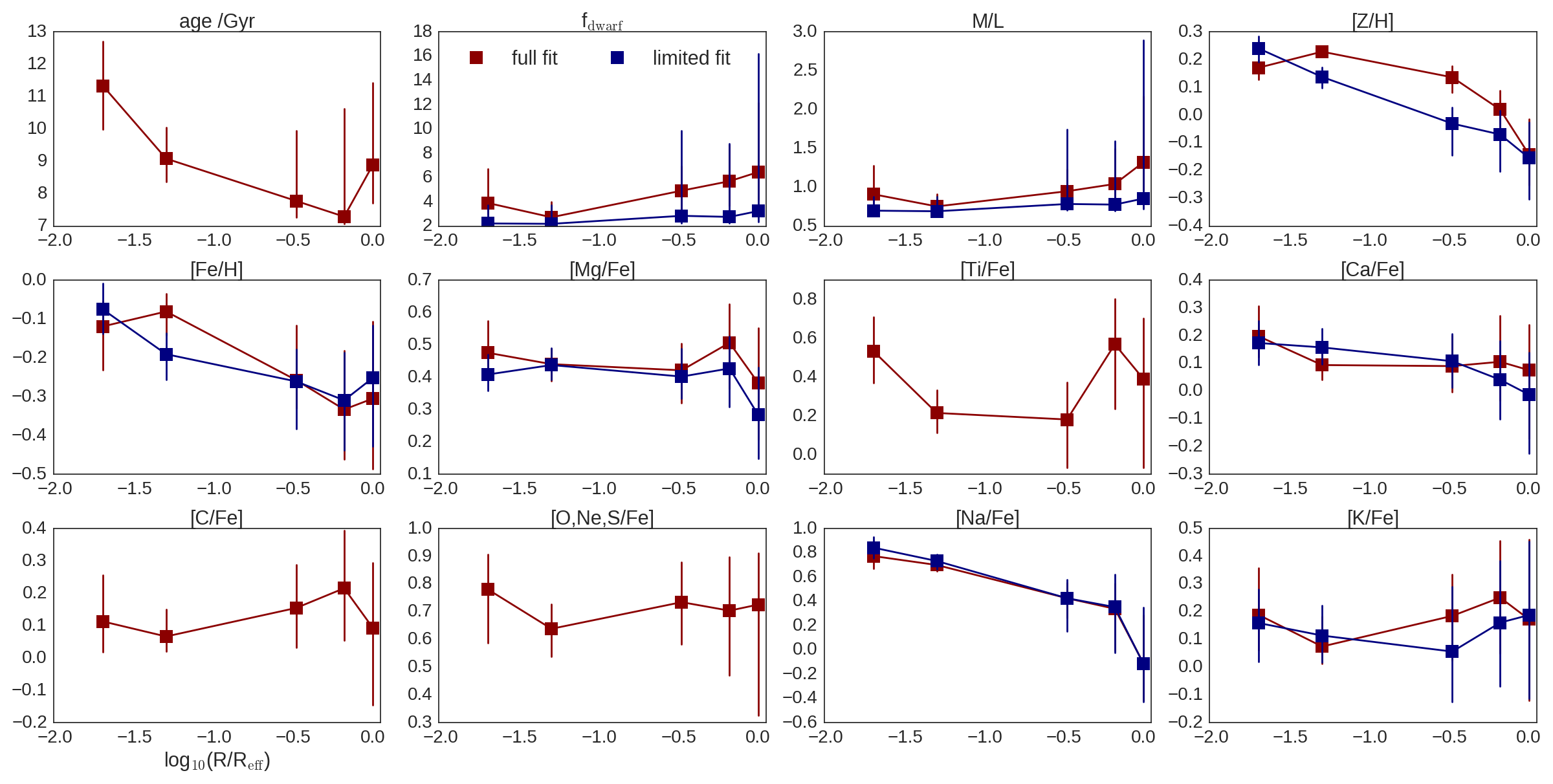}
		\caption[radial variation of stellar population parameters]{Parameter radial variations via two model fits. The parameters inferred from the stacked spectra using only a limited set of indices and a restricted parameter set are shown in blue, while those inferred when data from the K-band and H$\beta$ and additional model parameters are included are shown in red. Generally, as expected, our results are not radically altered with respect to those obtained previously, although the best-fit [Z/H] is an exception.}
		\label{fig:inference_comparison}
	\end{figure}
\end{landscape}

Including the K-band measurements and additional model parameters modifies our results to some extent. In Fig. \ref{fig:inference_comparison} we show the parameters inferred from the stacks using only a limited set of indices (Mg\textit{b}, Fe\,5015, IZ- and YJ-band measurements) and parameters (blue) and those inferred when data from the K-band and H$\beta$ are included along with a comprehensive set of model parameters (red). Generally, as expected, our results are not radically altered with respect to those obtained previously, although the best-fit [Z/H] is an exception. The inferred IMF is consistent with Milky Way-like throughout. The inferred age is consistent being constant with radius (approx. 10\,Gyr) or else gradually declining with radius. While we initially assumed an extremely old stellar population this slight difference does not in practice affect the inferred values of the other model parameters.

\begin{landscape}
	\begin{figure}
		\centering
		\vskip 20mm
		\includegraphics[scale=0.47]{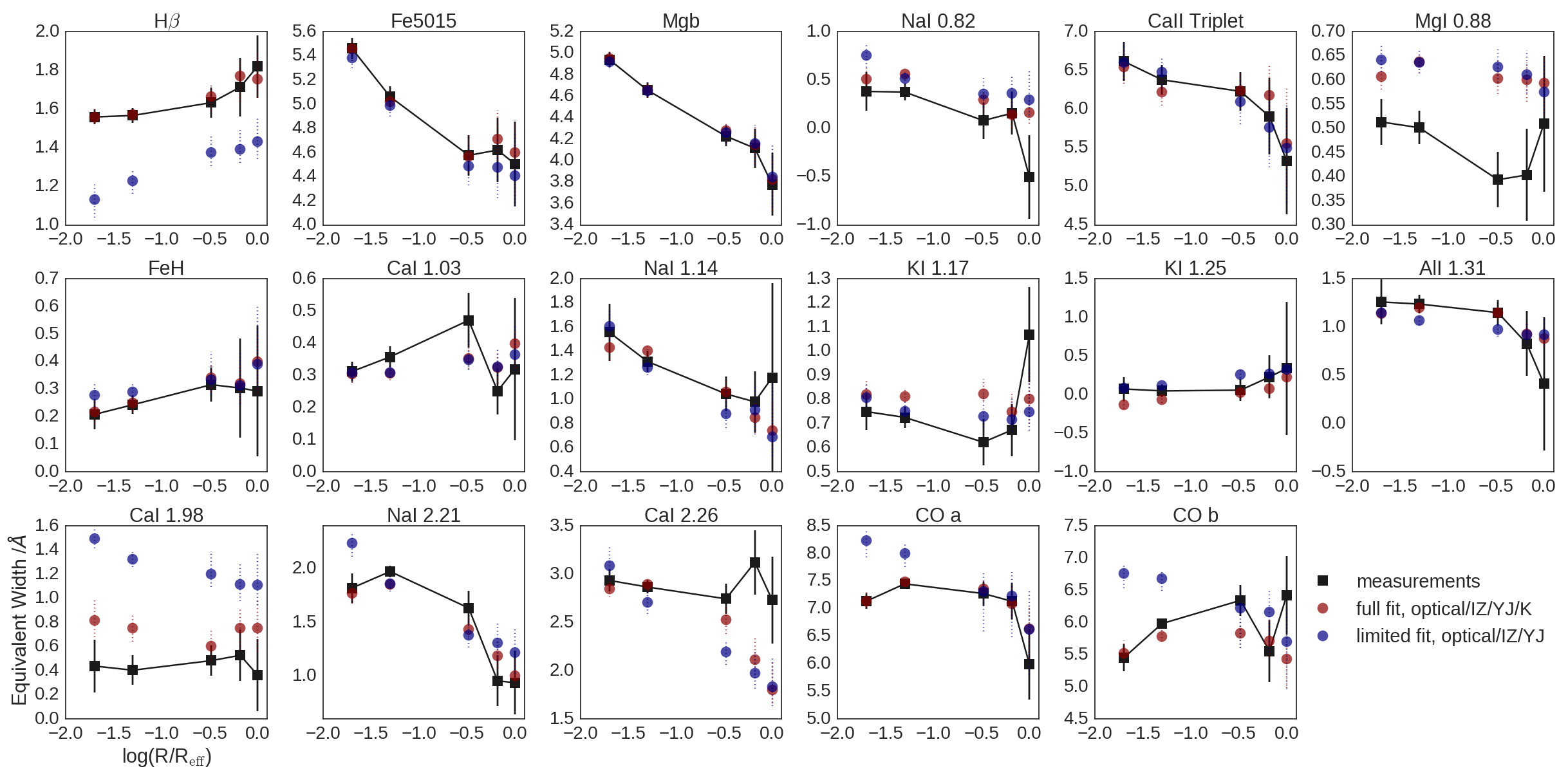}
		\caption[updated model predictions]{Two sets of model predictions (equivalent widths in \AA) generated from the parameters shown in Fig. \ref{fig:inference_comparison}. The simpler model (blue) was not fit to H$\beta$, Mg\,I 0.88$\upmu$m, Al\,I 1.31$\upmu$m, or any of the K-band indices. The more complex model uses more parameters and is fit to H$\beta$ and the K-band features (except CO $b$). As with the simpler model, it is not fit to  Mg\,I 0.88$\upmu$m or Al\,I 1.31$\upmu$m.}
		\label{fig:model_index_predictions}
	\end{figure}
\end{landscape}

In Fig. \ref{fig:model_index_predictions} we compare the two sets of model predictions to the actual measured line indices. This demonstrates several key points. For the most part, the simpler model reproduces the optical, IZ band, and YJ band indices adequately. However, as in Paper I it is challenging to simultaneously fit Na\,I 0.82\,$\upmu$m and Na\,I 1.14\,$\upmu$m. By contrast, the updated model now underpredicts the strength of H$\beta$, whereas previously the model tended to overpredict H$\beta$. As noted in Paper I, increased abundance of carbon can weaken H$\beta$. The inclusion of a metallicity parameter in the model means we now implicitly assume [C/H] rises in the core along with the $\alpha$-elements and Fe, explaining this change. Likewise, while the CvD16 model grids do not include variations in [Al/Fe], the Al\,I 1.31\,$\upmu$m feature's strength is no longer in strong tension with the model, suggesting that the assumption that aluminium tracks overall metallicity is reasonable. This is a significant advantage of the new model formalism. 

However, the simpler model does a poor job of predicting the K-band indices (we remind the reader that the simpler model was not fit to these indices). This shows that the K-band indices are adding useful information, which we take advantage of by using the more sophisticated model with additional parameters (a two-part power law IMF, stellar population age, [C/Fe], [Ti/Fe], and [O,\,Ne,\,S/Fe]). This model does a much better, though still imperfect job of fitting the K-band data, without compromising the fit to the bluer spectral features.

\begin{landscape}	
	\begin{figure}
		\centering
		\vskip 20mm
		\includegraphics[scale=0.47]{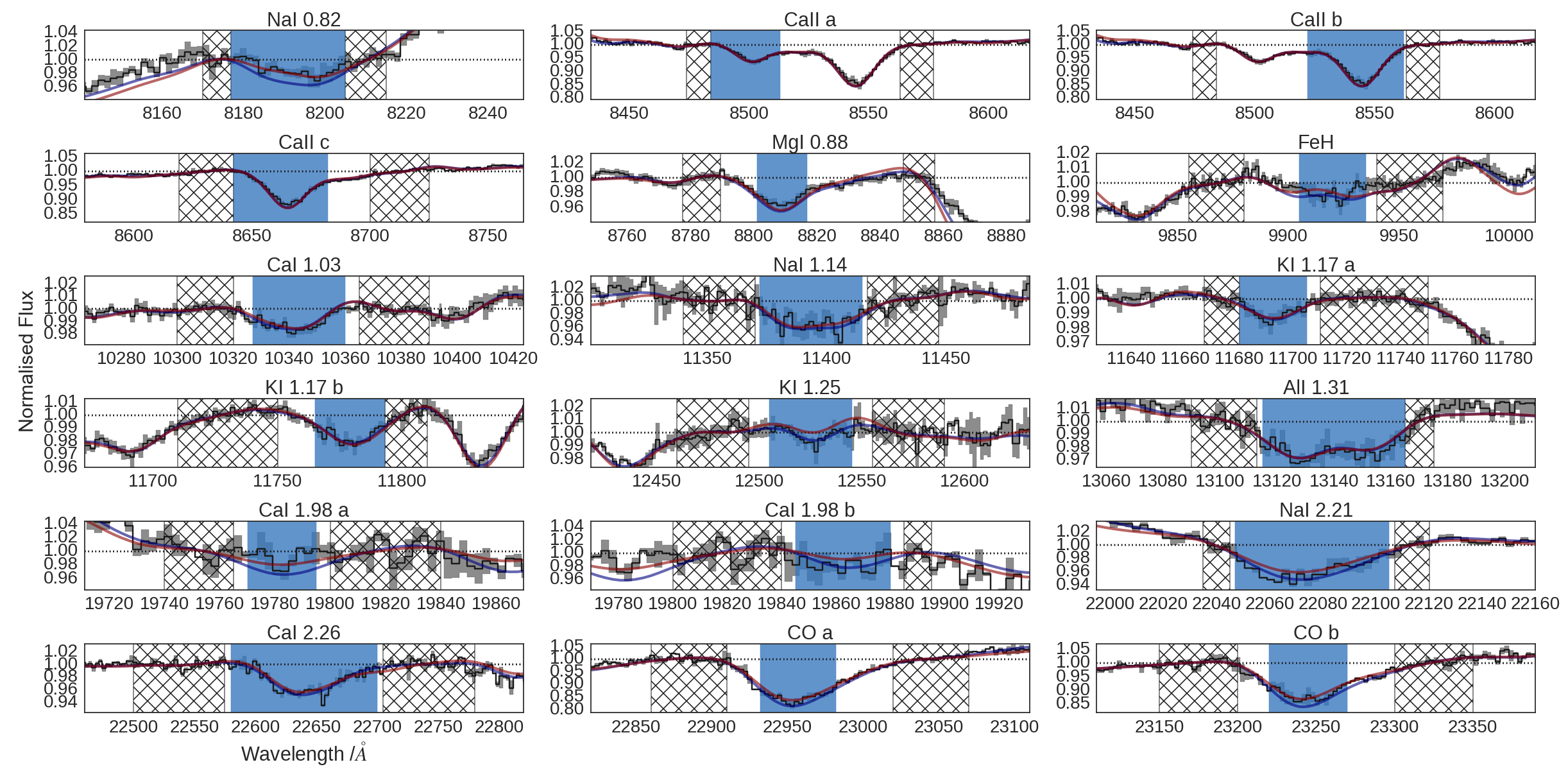}
		\caption[reconstructed model spectra, compared with the data]{The centrally extracted stacked spectrum in the region of each index definition with shaded uncertainties, alongside models generated from the parameters shown in Fig. \ref{fig:inference_comparison} (same colour scheme). The feature band for each index is shaded blue, while the pseudo-continuum regions are hatched. Although the model parameters are inferred from the measured equivalent widths, in most cases the models match the data within the uncertainties fairly well, especially for the strongest features. Note that spectra are normalised with respect to the pseudocontinuum (e.g. Na\,I 0.82$\upmu$m looks quite different to its appearance in the `raw' spectra because of this).}
		\label{fig:model_spectrum_predictions}
	\end{figure}
\end{landscape}
	
In Fig. \ref{fig:model_spectrum_predictions} we show how the CvD16 spectral models corresponding to these parameters appear when directly compared to the data. Specifically, we show the spectral region around each index for the centrally extracted stack, normalised by the pseudo-continuum bands on either side of each feature. We thus view the stacked spectrum as it appears when the index measurements are made. We remind the reader that the shaded regions of uncertainty correspond to the scatter amongst the spectra used to construct the stack, rather than the formal statistical error. We constructed the overplotted spectral models by linear interpolation of the CvD16 models, broadening this model to the velocity dispersion found using pPXF. We can thereby closely examine those features where the measured index strength is less well-fit by the model and perhaps determine the cause.
	
\subsection{Results for individual galaxies}

\begin{figure*}
	\centering
	\includegraphics[width=0.99\textwidth]{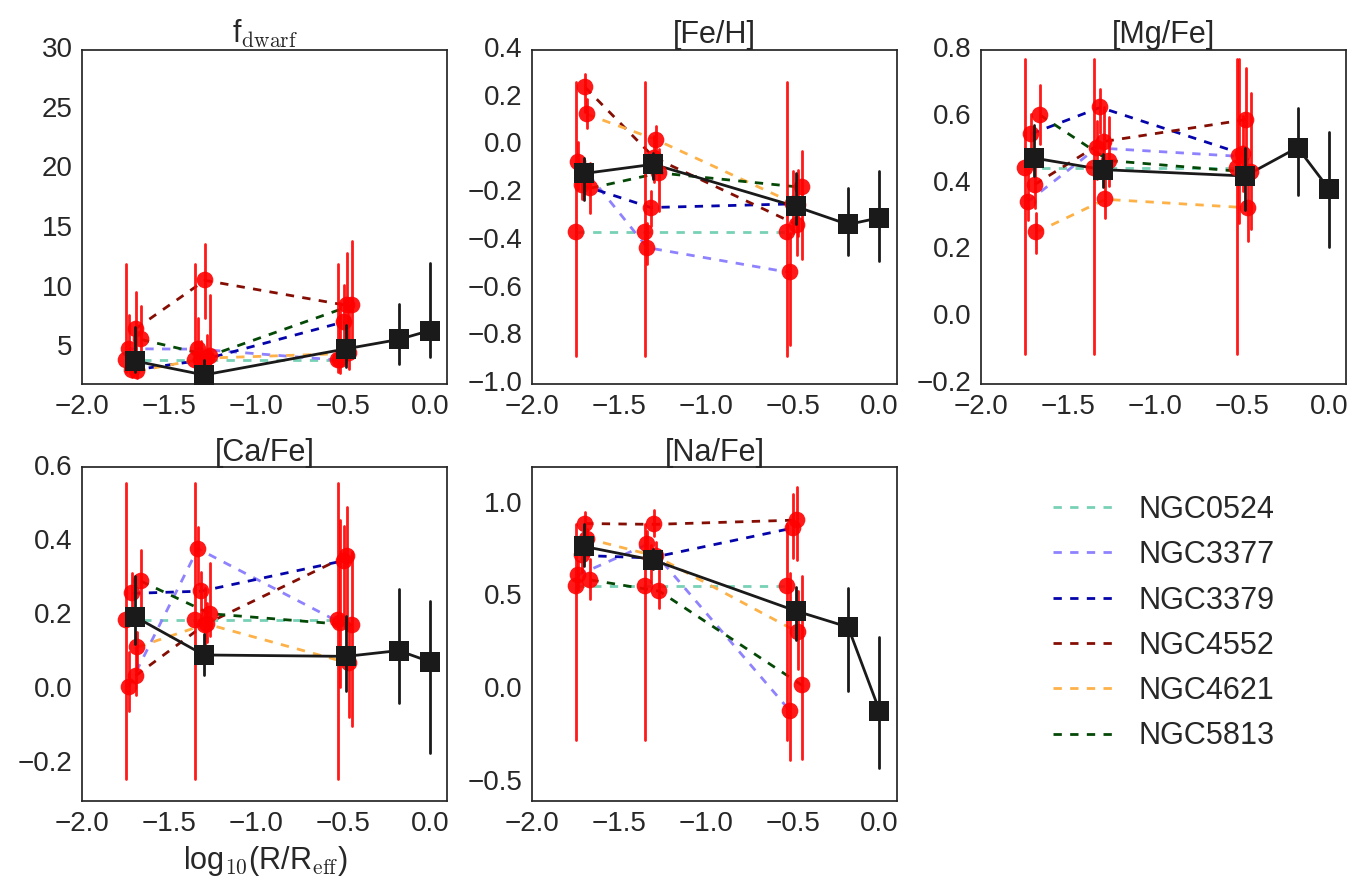}
	\caption[parameter values for individual galaxies]{Stellar population fitting results for individual galaxies (red circles) as well as the stack (black squares). The fit incorporates the K-band data where available and shows the behaviour of key (well-constrained) parameters. We exclude NGC\,1407 and NGC\,4486 as without good-quality optical data the fits are not well-constrained.}
	\label{fig:full_SP_results}
\end{figure*}

In this section we report the results for individual galaxies, comparing these with the average radial trends obtained from the stacked data. We fit these galaxies with the same set of model parameters and features as the stack where possible, although for some galaxies a full set of features is not available (note that we do not provide fits for NGC\,4486 since, as noted in Paper I, the optical data suffer from contamination by AGN emission). Results are provided for the three innermost extraction regions (the subdivided central IFU field and the innermost ring of IFUs at $\sim\frac{1}{3}$\,R$_{\rm eff}$) where the S/N is sufficient. For each extraction region we present a table of results spanning the full sample (see Tables \ref{table:R0_results}, \ref{table:R1_results}, and \ref{table:R2_results}). Note that we did not obtain additional K-band data for NGC\,0524 but have nonetheless fit the data via the updated model formalism. Likewise, we have created fits for NGC\,1407 and NGC\,3379 for which YJ-band data were not available. These results are presented in graphical form for key parameters in Fig. \ref{fig:full_SP_results}.
	
\begin{table*}
	\centering
	\caption[parameter estimates for individual galaxies (R\,$\sim$0.02\,R$_{\rm eff}$)]{Estimated parameter values for the spectra extracted at $\sim$\,0.02\,R$_{\rm eff}$, acquired using the updated models and the full set of index measurements. We fit for population age, a two-part power law IMF, and [Z/H], as well as the abundances of various elements of interest. Note that we did not obtain additional K-band data for NGC\,0524 but have nonetheless fit the data via the updated model formalism.}
	\begin{tabular}{lcccccccc}
		\hline
		                          &         NGC\,0524         &         NGC\,1407         &         NGC\,3377         &         NGC\,3379         &         NGC\,4552         &         NGC\,4621          &         NGC\,5813         &  \\ \hline
		age /Gyr                  & \gp{}10.2\err{2.40}{1.96} & \gp{}7.36\err{2.85}{0.25} & \gp{}7.08\err{0.58}{0.06} & \gp{}7.11\err{0.89}{0.07} & \gp{}7.09\err{0.62}{0.06} & \gp{}12.69\err{0.87}{2.33} & \gp{}8.04\err{2.24}{0.73} &  \\
		X1                        & \gp{}2.06\err{0.73}{0.70} & \gp{}1.83\err{0.59}{0.54} & \gp{}2.68\err{0.38}{1.02} & \gp{}1.10\err{0.89}{0.07} & \gp{}2.47\err{0.55}{0.92} & \gp{}1.24\err{0.66}{0.16}  & \gp{}1.56\err{0.71}{0.38} &  \\
		X2                        & \gp{}1.46\err{1.10}{0.29} & \gp{}1.16\err{1.31}{0.11} & \gp{}1.10\err{0.74}{0.07} & \gp{}1.22\err{0.44}{0.15} & \gp{}1.16\err{1.14}{0.11} & \gp{}1.08\err{0.53}{0.05}  & \gp{}2.04\err{0.87}{0.66} &  \\
		$\rm [Z/H]$               & \gp{}0.09\err{0.06}{0.08} & \gp{}0.17\err{0.10}{0.17} & \gp{}0.15\err{0.04}{0.04} & \gp{}0.24\err{0.06}{0.05} & \gp{}0.28\err{0.04}{0.05} & \gp{}0.22\err{0.05}{0.05}  & \gp{}0.13\err{0.06}{0.05} &  \\
		$\rm [Fe/H]$              & \gp{}0.04\err{0.09}{0.17} &  --0.21\err{0.17}{0.13}   &  --0.07\err{0.08}{0.11}   &  --0.17\err{0.06}{0.06}   & \gp{}0.24\err{0.05}{0.10} & \gp{}0.13\err{0.06}{0.06}  &  --0.18\err{0.11}{0.10}   &  \\
		$\rm [Mg/Fe]$             & \gp{}0.38\err{0.12}{0.08} & \gp{}0.79\err{0.14}{0.34} & \gp{}0.34\err{0.10}{0.05} & \gp{}0.55\err{0.06}{0.05} & \gp{}0.40\err{0.06}{0.07} & \gp{}0.25\err{0.06}{0.06}  & \gp{}0.60\err{0.09}{0.09} &  \\
		$\rm [Na/Fe]$             & \gp{}0.35\err{0.09}{0.09} & \gp{}0.71\err{0.15}{0.30} & \gp{}0.62\err{0.08}{0.08} & \gp{}0.73\err{0.07}{0.09} & \gp{}0.90\err{0.06}{0.06} & \gp{}0.82\err{0.05}{0.05}  & \gp{}0.59\err{0.11}{0.10} &  \\
		$\rm [Ca/Fe]$             & \gp{}0.13\err{0.14}{0.07} & \gp{}0.36\err{0.15}{0.12} & \gp{}0.01\err{0.09}{0.06} & \gp{}0.26\err{0.05}{0.06} & \gp{}0.04\err{0.08}{0.05} & \gp{}0.12\err{0.04}{0.06}  & \gp{}0.29\err{0.08}{0.10} &  \\
		$\rm [O,\,Ne,\,S/Fe]$     &  --0.24\err{0.22}{0.19}   &  --0.05\err{0.63}{0.19}   & \gp{}0.55\err{0.16}{0.12} & \gp{}0.89\err{0.10}{0.17} &  --0.13\err{0.16}{0.22}   & \gp{}0.61\err{0.07}{0.09}  & \gp{}0.53\err{0.22}{0.20} &  \\
		$\rm [Ti/Fe]$             & \gp{}0.39\err{0.24}{0.18} & \gp{}0.84\err{0.13}{0.43} & \gp{}0.40\err{0.19}{0.14} & \gp{}0.64\err{0.12}{0.11} & \gp{}0.23\err{0.15}{0.14} &   --0.23\err{0.13}{0.09}   & \gp{}0.71\err{0.16}{0.20} &  \\
		$\rm [C/Fe]$              & \gp{}0.16\err{0.15}{0.22} & \gp{}0.20\err{0.18}{0.15} & \gp{}0.36\err{0.12}{0.08} & \gp{}0.61\err{0.06}{0.06} & \gp{}0.16\err{0.08}{0.10} & \gp{}0.12\err{0.08}{0.09}  & \gp{}0.14\err{0.15}{0.12} &  \\
		$\rm [K/Fe]$              & \gp{}0.17\err{0.17}{0.19} & \gp{}0.55\err{0.11}{0.35} & \gp{}0.15\err{0.16}{0.15} & \gp{}0.36\err{0.22}{0.15} &  --0.05\err{0.18}{0.13}   & \gp{}0.09\err{0.13}{0.16}  & \gp{}0.13\err{0.23}{0.12} &  \\ \hline
		f$_{\mathrm{dwarf}}$ (\%) & \gp{}4.83\err{4.97}{1.60} & \gp{}4.41\err{3.60}{1.06} & \gp{}4.92\err{2.87}{1.57} & \gp{}3.12\err{1.53}{0.56} & \gp{}6.61\err{3.11}{2.32} & \gp{}3.06\err{1.29}{0.55}  & \gp{}5.71\err{2.81}{1.74} &  \\
		M/L                       & \gp{}1.04\err{0.71}{0.22} & \gp{}1.00\err{0.45}{0.15} & \gp{}1.00\err{0.48}{0.18} & \gp{}0.80\err{0.21}{0.07} & \gp{}1.08\err{0.58}{0.21} & \gp{}0.79\err{0.17}{0.06}  & \gp{}1.11\err{0.34}{0.21} &  \\ \hline
	\end{tabular}
 	\label{table:R0_results}
 \end{table*}
 
 \begin{table*}
 	\centering
 	\caption[parameter estimates for individual galaxies (R\,$\sim$\,0.05\,R$_{\rm eff}$)]{Estimated parameter values for the spectra extracted at $\sim$\,0.05\,R$_{\rm eff}$, acquired using the updated models and the full set of index measurements. We fit for population age, a two-part power law IMF, and [Z/H], as well as the abundances of various elements of interest. Note that we did not obtain additional K-band data for NGC\,0524 but have nonetheless fit the data via the updated model formalism.}
 	\begin{tabular}{lccccccc}
 		\hline
 		                          &         NGC\,0524          &         NGC\,1407         &         NGC\,3377         &         NGC\,3379         &         NGC\,4552          &         NGC\,4621          &         NGC\,5813         \\ \hline
 		age /Gyr                  & \gp{}13.54\err{0.32}{3.72} & \gp{}8.92\err{2.84}{1.32} & \gp{}7.16\err{1.12}{0.11} & \gp{}7.26\err{1.90}{0.19} & \gp{}7.09\err{0.54}{0.06}  & \gp{}10.61\err{1.74}{2.04} & \gp{}7.48\err{3.70}{0.33} \\
 		X1                        & \gp{}2.27\err{0.53}{0.77}  & \gp{}1.50\err{1.06}{0.33} & \gp{}2.45\err{0.58}{0.64} & \gp{}1.10\err{1.01}{0.07} & \gp{}3.36\err{0.10}{1.09}  & \gp{}1.20\err{0.84}{0.14}  & \gp{}1.88\err{0.67}{0.57} \\
 		X2                        & \gp{}1.14\err{1.43}{0.09}  & \gp{}2.08\err{0.79}{0.69} & \gp{}1.09\err{0.49}{0.06} & \gp{}1.10\err{0.63}{0.07} & \gp{}2.14\err{0.62}{0.60}  & \gp{}1.11\err{0.97}{0.08}  & \gp{}3.29\err{0.14}{1.63} \\
 		$\rm [Z/H]$               &   --0.01\err{0.06}{0.12}   &  --0.21\err{0.26}{0.16}   & \gp{}0.01\err{0.04}{0.05} & \gp{}0.11\err{0.07}{0.06} & \gp{}0.23\err{0.05}{0.05}  & \gp{}0.17\err{0.06}{0.05}  & \gp{}0.08\err{0.08}{0.08} \\
 		$\rm [Fe/H]$              &   --0.13\err{0.09}{0.13}   &  --0.32\err{0.18}{0.24}   &  --0.43\err{0.10}{0.07}   &  --0.26\err{0.07}{0.08}   &   --0.05\err{0.09}{0.10}   & \gp{}0.02\err{0.06}{0.09}  &  --0.12\err{0.10}{0.16}   \\
 		$\rm [Mg/Fe]$             & \gp{}0.42\err{0.12}{0.08}  & \gp{}0.49\err{0.28}{0.36} & \gp{}0.51\err{0.08}{0.09} & \gp{}0.63\err{0.05}{0.08} & \gp{}0.53\err{0.10}{0.07}  & \gp{}0.35\err{0.07}{0.06}  & \gp{}0.47\err{0.13}{0.08} \\
 		$\rm [Na/Fe]$             & \gp{}0.22\err{0.12}{0.11}  & \gp{}0.77\err{0.15}{0.26} & \gp{}0.79\err{0.07}{0.12} & \gp{}0.71\err{0.07}{0.13} & \gp{}0.89\err{0.08}{0.06}  & \gp{}0.72\err{0.08}{0.05}  & \gp{}0.53\err{0.11}{0.09} \\
 		$\rm [Ca/Fe]$             & \gp{}0.15\err{0.12}{0.06}  & \gp{}0.45\err{0.15}{0.17} & \gp{}0.38\err{0.06}{0.09} & \gp{}0.27\err{0.05}{0.08} & \gp{}0.18\err{0.10}{0.07}  & \gp{}0.18\err{0.06}{0.05}  & \gp{}0.21\err{0.14}{0.06} \\
 		$\rm [O,\,Ne,\,S/Fe]$     & \gp{}0.31\err{0.22}{0.25}  & \gp{}0.43\err{0.30}{0.43} & \gp{}0.69\err{0.19}{0.15} & \gp{}0.80\err{0.14}{0.31} & \gp{}0.10\err{0.14}{0.23}  & \gp{}0.68\err{0.09}{0.11}  & \gp{}0.41\err{0.25}{0.19} \\
 		$\rm [Ti/Fe]$             & \gp{}0.33\err{0.24}{0.25}  & \gp{}0.35\err{0.39}{0.33} & \gp{}0.79\err{0.18}{0.20} & \gp{}0.71\err{0.14}{0.15} & \gp{}0.64\err{0.16}{0.15}  &   --0.11\err{0.17}{0.10}   & \gp{}0.31\err{0.31}{0.18} \\
 		$\rm [C/Fe]$              & \gp{}0.20\err{0.14}{0.23}  & \gp{}0.16\err{0.28}{0.16} & \gp{}0.50\err{0.11}{0.11} & \gp{}0.58\err{0.08}{0.10} & \gp{}0.36\err{0.09}{0.14}  & \gp{}0.24\err{0.09}{0.11}  & \gp{}0.14\err{0.16}{0.15} \\
 		$\rm [K/Fe]$              & \gp{}0.15\err{0.16}{0.17}  & \gp{}0.25\err{0.25}{0.26} & \gp{}0.67\err{0.06}{0.17} & \gp{}0.44\err{0.16}{0.22} & \gp{}0.26\err{0.14}{0.15}  & \gp{}0.09\err{0.12}{0.13}  &  --0.05\err{0.16}{0.08}   \\ \hline
 		f$_{\mathrm{dwarf}}$ (\%) & \gp{}6.58\err{3.72}{2.34}  & \gp{}5.23\err{4.32}{1.63} & \gp{}4.89\err{2.59}{1.36} & \gp{}3.96\err{1.62}{1.09} & \gp{}10.66\err{3.04}{3.16} & \gp{}4.14\err{1.93}{1.06}  & \gp{}4.35\err{5.04}{1.13} \\
 		M/L                       & \gp{}1.31\err{0.50}{0.35}  & \gp{}1.07\err{0.60}{0.21} & \gp{}1.04\err{0.44}{0.19} & \gp{}0.89\err{0.24}{0.13} & \gp{}1.93\err{0.55}{0.55}  & \gp{}0.94\err{0.24}{0.14}  & \gp{}0.93\err{0.66}{0.13} \\ \hline
 	\end{tabular}
\label{table:R1_results}
\end{table*}

 \begin{table*}
 	\centering
 	\caption[parameter estimates for individual galaxies (R\,$\sim \frac{1}{3}$R$_{\rm eff}$)]{Estimated parameter values for the spectra extracted at $\sim \frac{1}{3}$R$_{\rm eff}$, acquired using the updated models and the full set of index measurements. We fit for population age, a two-part power law IMF, and [Z/H], as well as the abundances of various elements of interest. Note that we did not obtain additional K-band data for NGC\,0524 but have nonetheless fit the data via the updated model formalism.}
 	\begin{tabular}{lccccccc}
 		\hline
 		                          &         NGC\,0524          &         NGC\,1407         &         NGC\,3377         &         NGC\,3379         &         NGC\,4552         &         NGC\,4621          &         NGC\,5813         \\ \hline
 		age /Gyr                  & \gp{}11.18\err{1.80}{2.54} & \gp{}7.45\err{3.34}{0.29} & \gp{}9.14\err{2.86}{1.44} & \gp{}8.89\err{1.98}{1.22} & \gp{}8.56\err{3.23}{0.99} & \gp{}10.04\err{2.29}{1.84} & \gp{}9.69\err{2.93}{1.74} \\
 		X1                        & \gp{}2.37\err{0.67}{0.88}  & \gp{}1.66\err{0.80}{0.45} & \gp{}1.73\err{0.88}{0.48} & \gp{}2.12\err{0.63}{0.76} & \gp{}2.50\err{0.59}{0.89} & \gp{}1.86\err{0.81}{0.57}  & \gp{}3.05\err{0.28}{1.23} \\
 		X2                        & \gp{}2.61\err{0.55}{1.07}  & \gp{}1.12\err{1.10}{0.08} & \gp{}1.26\err{1.19}{0.17} & \gp{}2.10\err{0.81}{0.68} & \gp{}3.30\err{0.13}{1.40} & \gp{}1.31\err{0.96}{0.21}  & \gp{}1.22\err{1.36}{0.15} \\
 		$\rm [Z/H]$               &   --0.08\err{0.11}{0.14}   & \gp{}0.36\err{0.03}{0.25} &  --0.43\err{0.14}{0.27}   & \gp{}0.03\err{0.10}{0.09} & \gp{}0.02\err{0.09}{0.12} & \gp{}0.07\err{0.09}{0.10}  &  --0.15\err{0.16}{0.18}   \\
 		$\rm [Fe/H]$              & \gp{}0.05\err{0.10}{0.17}  &  --0.06\err{0.17}{0.16}   &  --0.53\err{0.16}{0.30}   &  --0.25\err{0.14}{0.12}   &  --0.33\err{0.21}{0.13}   &   --0.25\err{0.15}{0.13}   &  --0.18\err{0.15}{0.30}   \\
 		$\rm [Mg/Fe]$             & \gp{}0.18\err{0.17}{0.09}  & \gp{}0.59\err{0.26}{0.31} & \gp{}0.48\err{0.29}{0.20} & \gp{}0.49\err{0.12}{0.11} & \gp{}0.59\err{0.16}{0.15} & \gp{}0.33\err{0.15}{0.10}  & \gp{}0.43\err{0.24}{0.17} \\
 		$\rm [Na/Fe]$             &   --0.16\err{0.25}{0.27}   &  --0.07\err{0.17}{0.24}   &  --0.11\err{0.74}{0.27}   & \gp{}0.87\err{0.19}{0.16} & \gp{}0.92\err{0.18}{0.21} & \gp{}0.31\err{0.22}{0.20}  & \gp{}0.03\err{0.59}{0.40} \\
 		$\rm [Ca/Fe]$             &   --0.12\err{0.17}{0.10}   & \gp{}0.08\err{0.20}{0.15} & \gp{}0.18\err{0.28}{0.17} & \gp{}0.35\err{0.09}{0.15} & \gp{}0.36\err{0.13}{0.18} & \gp{}0.07\err{0.11}{0.15}  & \gp{}0.17\err{0.20}{0.27} \\
 		$\rm [O,\,Ne,\,S/Fe]$     &   --0.24\err{0.32}{0.28}   & \gp{}0.61\err{0.25}{0.47} & \gp{}0.25\err{0.47}{0.34} & \gp{}0.56\err{0.28}{0.52} & \gp{}0.29\err{0.33}{0.26} & \gp{}0.81\err{0.15}{0.18}  & \gp{}0.15\err{0.51}{0.43} \\
 		$\rm [Ti/Fe]$             &   --0.09\err{0.36}{0.24}   & \gp{}0.67\err{0.18}{0.39} & \gp{}0.51\err{0.37}{0.32} & \gp{}0.23\err{0.33}{0.26} & \gp{}0.64\err{0.22}{0.33} & \gp{}0.02\err{0.31}{0.16}  & \gp{}0.22\err{0.46}{0.37} \\
 		$\rm [C/Fe]$              &   --0.16\err{0.24}{0.18}   & \gp{}0.17\err{0.20}{0.19} & \gp{}0.14\err{0.28}{0.31} & \gp{}0.26\err{0.18}{0.15} & \gp{}0.24\err{0.18}{0.19} & \gp{}0.22\err{0.16}{0.16}  & \gp{}0.03\err{0.33}{0.22} \\
 		$\rm [K/Fe]$              &   --0.35\err{0.32}{0.10}   & \gp{}0.15\err{0.31}{0.21} & \gp{}0.05\err{0.31}{0.26} & \gp{}0.24\err{0.25}{0.21} & \gp{}0.16\err{0.26}{0.20} & \gp{}0.19\err{0.28}{0.17}  & \gp{}0.21\err{0.21}{0.31} \\ \hline
 		f$_{\mathrm{dwarf}}$ (\%) & \gp{}7.00\err{5.31}{2.63}  & \gp{}4.47\err{3.52}{1.43} & \gp{}3.96\err{4.84}{1.04} & \gp{}7.18\err{3.10}{2.86} & \gp{}8.56\err{4.43}{3.00} & \gp{}4.53\err{3.75}{1.29}  & \gp{}8.59\err{5.40}{4.16} \\
 		M/L                       & \gp{}1.27\err{0.84}{0.33}  & \gp{}0.93\err{0.49}{0.15} & \gp{}0.89\err{0.65}{0.12} & \gp{}1.35\err{0.47}{0.40} & \gp{}1.34\err{0.77}{0.31} & \gp{}0.99\err{0.53}{0.17}  & \gp{}0.97\err{1.03}{0.17} \\ \hline
 	\end{tabular}
\label{table:R2_results}
\end{table*}

\subsection{Analysis of results}

\subsubsection*{[O,\,Ne,\,S/Fe]}

In the full fit, we marginalise over the nuisance parameter [O,\,Ne,\,S/Z] which captures the (small) model variations due to variations in poorly-constrained $\alpha$-elements. As with other abundance parameters, we convert this to the more informative form [O,\,Ne,\,S/Fe]. As is clear from Fig. \ref{fig:inference_comparison}, the inferred values of [O,\,Ne,\,S/Fe] are extremely high, albeit poorly constrained. While the expected enhancement should be similar to that of [Mg/Fe], the most probable values appear to be systematically offset by approximately 1$\sigma$. This appears to be driven by the Na\,I 0.82\,$\upmu$m feature, which can be weakened somewhat by large values of this parameter, reducing the tension with the redder Na\,I feature strengths. In Table \ref{table:noO_results} we present the results of fitting the stacks without this parameter (assuming [O,\,Ne,\,S/Fe] = 0 instead). This allows us to show that the principal knock-on effect of including the [O,\,Ne,\,S/Fe] parameter is to increase the inferred [Na/Fe] value, since larger values can be reconciled with the Na\,I 0.82\,$\upmu$m feature's strength. The effect on the IMF is negligible.

\begin{table*}
	\centering
	\caption[parameter values estimated when not accounting for abundance variations in O, Ne, and S.]{Estimated parameter values for the stacked spectra, acquired using the updated models and the full set of index measurements. In this table the nuisance parameter [O,\,Ne,\,S/Fe] has been excluded from the fit. R1 and R2 correspond to the two extraction regions in the central IFU ($<0.7\arcsec$, $>0.7\arcsec$) while R3, R4, and R5 correspond to the rings of IFUs arranged at $\sim1/3\,$R$_{\rm eff}$, $\sim2/3\,$R$_{\rm eff}$, and $\sim$\,R$_{\rm eff}$.}
	\begin{tabular}{lcccccc}
		\hline
		&            R1             &            R2             &            R3             &            R4             &            R5             &      gradient      \\ \hline
		age /Gyr             & \gp{}8.73\rerr{0.88}{1.01} & \gp{}7.34\rerr{0.23}{0.89} & \gp{}7.17\rerr{0.12}{1.32} & \gp{}7.38\rerr{0.25}{2.33} & \gp{}7.88\rerr{0.59}{2.35} &  --0.71$\pm$0.55   \\
		X1                   & \gp{}1.79\rerr{0.51}{0.70} & \gp{}1.81\rerr{0.56}{0.49} & \gp{}1.60\rerr{0.38}{0.70} & \gp{}1.13\rerr{0.09}{1.14} & \gp{}1.32\rerr{0.22}{1.40} &  --0.32$\pm$0.23   \\
		X2                   & \gp{}1.19\rerr{0.13}{0.87} & \gp{}1.08\rerr{0.05}{0.51} & \gp{}1.16\rerr{0.11}{0.61} & \gp{}1.17\rerr{0.11}{0.98} & \gp{}1.58\rerr{0.37}{1.07} & \gp{}0.14$\pm$0.14 \\
		$\rm [Z/H]$          & \gp{}0.17\rerr{0.03}{0.03} & \gp{}0.24\rerr{0.03}{0.02} & \gp{}0.15\rerr{0.04}{0.03} & \gp{}0.03\rerr{0.09}{0.06} &  --0.14\rerr{0.10}{0.11}   &  --0.09$\pm$0.03   \\
		$\rm [Fe/H]$         &  --0.04\rerr{0.07}{0.05}   &  --0.07\rerr{0.06}{0.04}   &  --0.26\rerr{0.08}{0.11}   &   --0.3\rerr{0.11}{0.15}   &  --0.19\rerr{0.21}{0.17}   &  --0.16$\pm$0.05   \\
		$\rm [Mg/Fe]$        & \gp{}0.49\rerr{0.04}{0.08} & \gp{}0.48\rerr{0.04}{0.05} & \gp{}0.48\rerr{0.08}{0.08} & \gp{}0.50\rerr{0.13}{0.11} & \gp{}0.30\rerr{0.16}{0.21} &  --0.02$\pm$0.04   \\
		$\rm [Na/Fe]$        & \gp{}0.63\rerr{0.11}{0.09} & \gp{}0.64\rerr{0.05}{0.05} & \gp{}0.32\rerr{0.15}{0.13} & \gp{}0.06\rerr{0.23}{0.26} &  --0.24\rerr{0.27}{0.40}   &  --0.37$\pm$0.09   \\
		$\rm [Ca/Fe]$        & \gp{}0.13\rerr{0.06}{0.08} & \gp{}0.06\rerr{0.05}{0.06} & \gp{}0.10\rerr{0.10}{0.08} & \gp{}0.10\rerr{0.17}{0.11} &  --0.12\rerr{0.18}{0.20}   &  --0.05$\pm$0.05   \\
		$\rm [C/Fe]$         & \gp{}0.06\rerr{0.09}{0.09} & \gp{}0.11\rerr{0.06}{0.06} & \gp{}0.21\rerr{0.12}{0.08} & \gp{}0.17\rerr{0.15}{0.14} &  --0.01\rerr{0.24}{0.21}   & \gp{}0.06$\pm$0.05 \\
		$\rm [Ti/Fe]$        & \gp{}0.52\rerr{0.13}{0.13} & \gp{}0.36\rerr{0.1}{0.09}  & \gp{}0.37\rerr{0.21}{0.17} & \gp{}0.62\rerr{0.34}{0.19} & \gp{}0.48\rerr{0.55}{0.30} &  --0.02$\pm$0.11   \\
		$\rm [K/Fe]$         & \gp{}0.06\rerr{0.10}{0.21} & \gp{}0.06\rerr{0.05}{0.10} & \gp{}0.2\rerr{0.14}{0.21}  & \gp{}0.20\rerr{0.20}{0.26} & \gp{}0.13\rerr{0.26}{0.27} & \gp{}0.09$\pm$0.07 \\ \hline
		f$_{\mathrm{dwarf}}$ & \gp{}4.2\rerr{1.1}{2.7}\%  & \gp{}3.0\rerr{0.6}{1.7}\%  & \gp{}4.0\rerr{1.1}{1.8}\%  & \gp{}4.3\rerr{1.3}{3.1}\%  & \gp{}6.3\rerr{2.5}{4.7}\%  & \gp{}0.7$\pm$1.3\% \\
		M/L                  & \gp{}0.89\rerr{0.11}{0.39} & \gp{}0.79\rerr{0.07}{0.23} & \gp{}0.83\rerr{0.08}{0.28} & \gp{}0.91\rerr{0.14}{0.45} & \gp{}0.96\rerr{0.17}{0.84} & \gp{}0.04$\pm$0.08 \\ \hline
	\end{tabular}
	\label{table:noO_results}
\end{table*}

\subsubsection*{Mg\,I 0.88\,$\upmu$m}

\begin{figure}
	\includegraphics[width=0.49\textwidth]{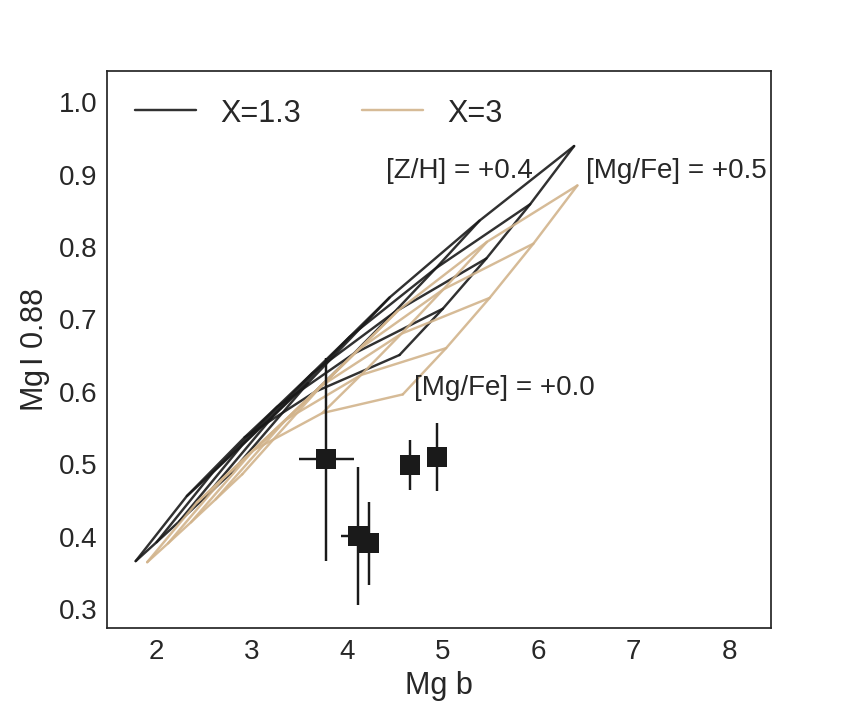}
	\caption[Mg\,\textit{b} vs. Mg\,I 0.88\umicron\, index-index grid, compared with the data]{Equivalent widths (\AA) of the Mg features, measured on the stacked spectra, along with model grids where the vertices indicate variation of metallicity and [Mg/Fe]. IMF variation makes relatively little difference to these features. The data and grids appear to be offset from each other.}
	\label{fig:mg_idx_idx}
\end{figure}

As mentioned above, in the updated model framework it is challenging to fit our measurements of the Mg\,I 0.88\,$\upmu$m feature, and so we did not include this feature in the fits so as to avoid biasing them. In Fig. \ref{fig:mg_idx_idx} we show an index-index plot for the stack measurements; we also show the model grids under variation of metallicity and [Mg/Fe] for a bottom-light and a bottom-heavy IMF. The data appear offset from the model grid.

An examination of the Mg\,I 0.88\,$\upmu$m line in Fig. \ref{fig:model_spectrum_predictions} suggests that the issue could be related to the pseudo-continuum definition in conjunction with differences between the model and the data in terms of long range spectral variations. On the red side of the feature the pseudo-continuum band could be affected by the Ti\,I 0.89\,$\upmu$m break.

\subsubsection*{[Na/Fe]}

As in Paper I, the recovered sodium abundance is often very high. Once again, there is tension in the best-fit model between the various Na\,I features. Examination of Fig. \ref{fig:model_index_predictions} suggests that the model tends to overestimate the strength of Na\,I 0.82\,$\upmu$m and underestimate the strength of Na\,I 2.21\,$\upmu$m.

\begin{figure*}
	\includegraphics[width=0.99\textwidth]{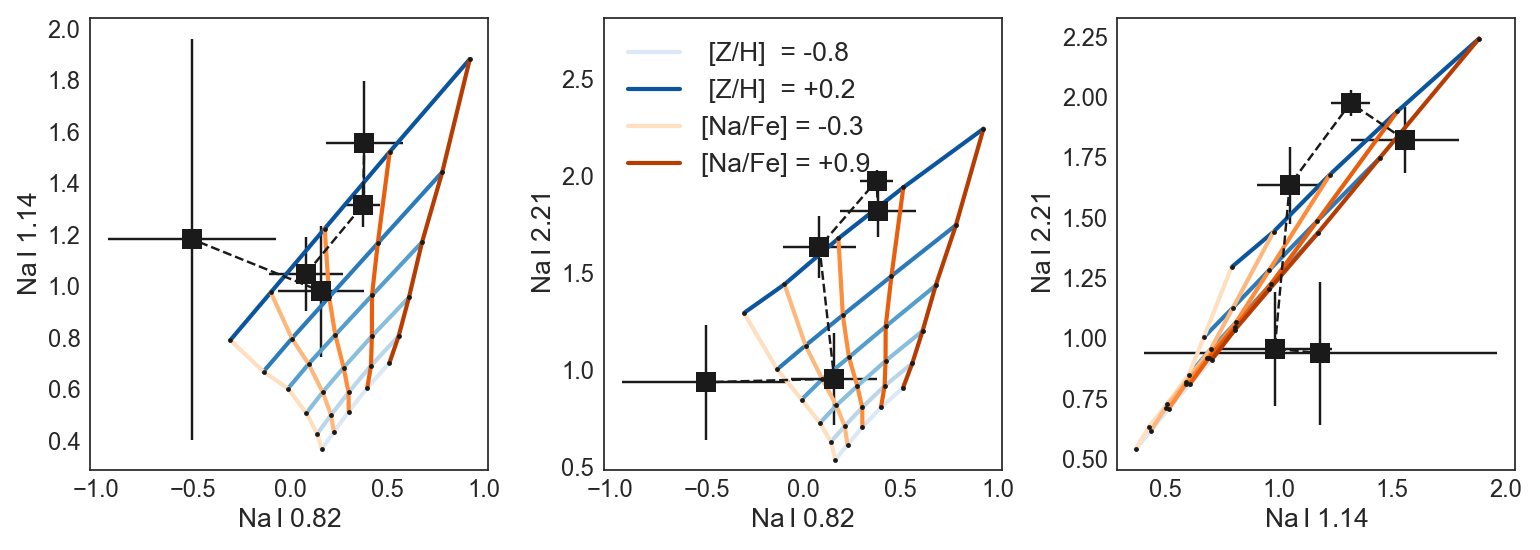}
	\caption[index-index grids for three Na features]{Equivalent widths of the Na features, measured on the stacked spectra, along with model grids where the vertices indicate variation of metallicity and [Na/Fe], assuming a Milky Way-like IMF. Blue lines are lines of constant [Z/H], orange lines are lines of constant [Na/Fe]. Darker colours indicate higher values. Finally, dashed lines link data points from adjacent radial extraction zones.}
	\label{fig:na_idx_idx}
\end{figure*}

As was discussed in \cite{2015MNRAS.454L..71S} the Na\,I 1.14\,$\upmu$m line has been found to be unusually strong in massive ETGs, which is difficult to reconcile with the SSP models without violating other constraints. Our finding of a discrepancy between the Na\,I 0.82\,$\upmu$m and Na\,I 2.21\,$\upmu$m lines indicates that the Na\,I 2.21\,$\upmu$m line is similarly (or perhaps even \textit{more}) discrepant. This contrasts with the findings of \cite{2017MNRAS.464.3597L}, in which two massive ETGs were studied. The authors found that a different set of stellar population models that include \textit{coupled} spectral responses to changes in the IMF and Na abundance are able to simultaneously match the strength of the Na\,I lines.

The discrepancy is shown particularly clearly in Fig. \ref{fig:na_idx_idx}, which shows three index-index plots for the stacked spectra. Na\,I 0.82\,$\upmu$m is comparatively insensitive to [Z/H] and is best-fit with a fairly low [Na/Fe] in all cases. By contrast, the other two lines require \textit{either} extremely high [Z/H] or [Na/Fe] (in the case of Na\,I 1.14\,$\upmu$m, the required [Z/H] would be prohibitively high). We may also examine the relevant panels in Fig. \ref{fig:model_spectrum_predictions}, which shows the features from the centrally-extracted stacked spectrum alongside the models. Na\,I 0.82\,$\upmu$m can be reproduced by the more complex model, but as just discussed is weaker than can be accounted for with the simpler model. Na\,I 1.14\,$\upmu$m is reproduced by both models, but Na\,2.21\,$\upmu$m appears discrepant in shape on the blue side of the feature.

\subsubsection*{The IMF: bottom-heavy or Milky Way-like?}

The CvD16 models allow the IMF additional flexibility, with the functional form modelled as a two-part broken power law with slope X1 below 0.5\,M$_\odot$ and X2 between 0.5--1.0\,M$_\odot$. In practice, this means that the relative contributions of the least massive stars and the $\sim$\,0.5\,M$_\odot$ stars can be varied separately -- in reality of course varying the upper slope X2 will increase/decrease the contribution of both, so the two parameters have some degeneracy. 

For the stacked spectra, the IMF we recover is consistent with the Milky Way IMF in all cases (in contrast to Paper I, where a modestly bottom-heavy IMF was preferred). The individual galaxies show scatter of $\sim$\,2\% in \fdwarf around a mean of 5.6\% (where a Kroupa IMF corresponds to \fdwarf\,=\,5\%) and in almost all cases are consistent with this mean value within the estimated statistical uncertainties. The individual galaxies appear offset from the stack in one case: this may be indicative of unaccounted-for systematics (the [Ca/Fe] abundance shows an offset at a similar level in the same bin).

Compared to the results presented in Paper I, however, the distribution of IMFs is considerably tighter than before and with the addition of new data and updated models we infer lower contributions to the light from dwarf stars (e.g. for the centrally extracted stack, 90\% of draws from the posterior probability distribution correspond to \fdwarf$<$\,8.4\%, corresponding to the Salpeter IMF, whereas in Paper I we found \fdwarf\,=\,9.4$\pm$2.1\%).

\begin{figure}
	\centering
	\includegraphics[width=0.48\textwidth]{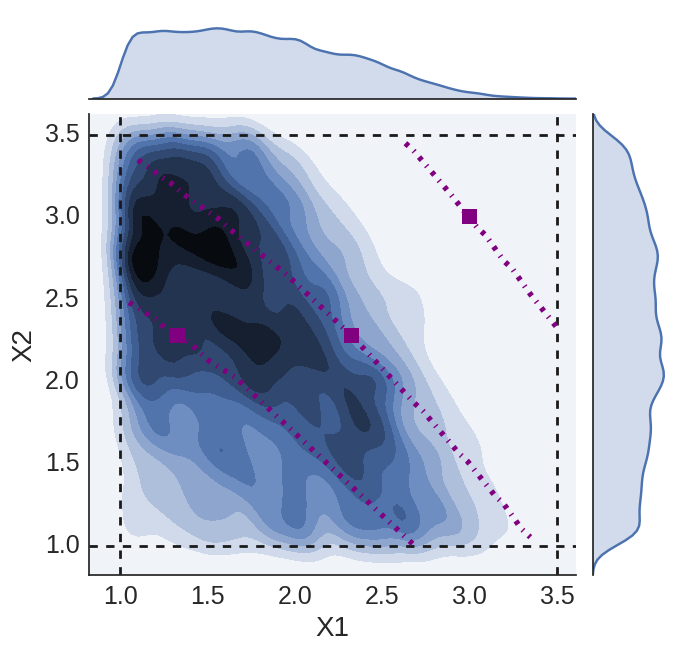}
	\caption[model covariance of flexible-IMF power law slopes]{The results of fitting feature strengths from the centrally extracted spectrum from NGC\,5813 with X1 and X2 varying freely. The joint and marginal distributions are shown via kernel-density estimation. Prior boundaries are indicated by the thick dashed lines, while dot-dashed contours indicate three lines of constant \fdwarf\,=\,5.0\%, 8.4\%, and 19.0\% (bottom-left to top-right). These values are equal to \fdwarf in the case of Kroupa, Salpeter, and X=3 IMFs respectively. Square markers show the location of these particular IMFs on each line.}
	\label{fig:X1X2_joint_example}
\end{figure}

The relationship between the IMF slope/slopes and the index strengths is highly non-linear. We now discuss the consequences of this for the inferred parameters. In Fig. \ref{fig:X1X2_joint_example} we give as an example the joint and marginal posterior probability distributions over X1 and X2 inferred for the innermost spectrum of NGC\,5813. The marginalised distributions of these two parameters are broad and somewhat skewed. While some covariance between the two parameters remains, this is mitigated by the use of multi-band data. To aid comparison, we plot three lines of constant f$_{\rm dwarf}$. While the form of the IMF varies between points on one of these lines, the chosen values of \fdwarf correspond to Milky Way-like, Salpeter, and X\,=\,3 power law IMFs respectively: most of the joint probability is enclosed between the first two of these.

\begin{figure}
	\centering
	\includegraphics[width=0.48\textwidth]{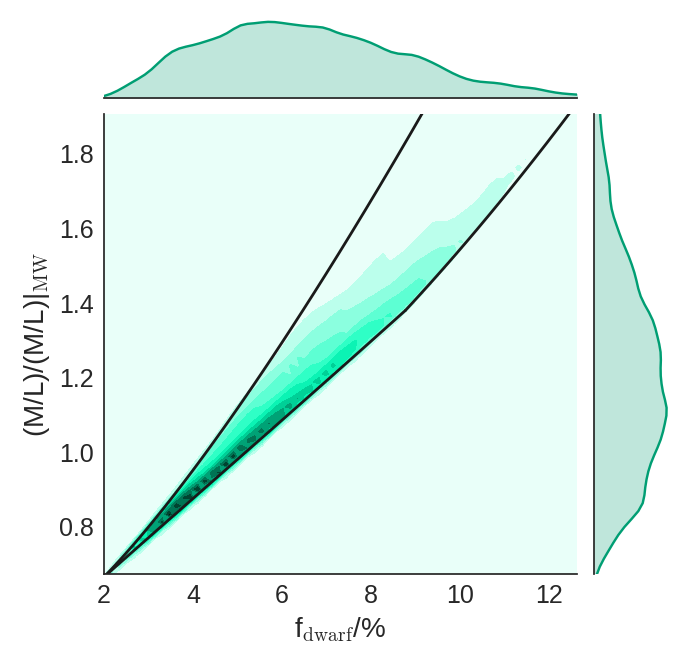}
	\caption[\fdwarf\, vs. M/L ratio]{This plot corresponds to Fig. \ref{fig:X1X2_joint_example}, with the inferred posterior distribution of X1 and X2 for NGC\,5813 recast into values of f$_{\rm dwarf}$ and V-band mass-to-light ratio (normalised so that M/L = 1 for a Milky Way-like population). The thick black line corresponds to the region enclosed by the priors on X1 and X2.}
	\label{fig:fd_ML_joint_example}
\end{figure}

In Fig. \ref{fig:fd_ML_joint_example} we transform this distribution into the posterior probability for f$_{\rm dwarf}$ and relative M/L. These parameters are quite strongly correlated, as is clear from the figure. Note that the posterior probability distributions are less skewed than for X1, X2 (as might be anticipated from the roughly linear dependence of f$_{\rm dwarf}$ on the index strengths, which have symmetric statistical uncertainties). Crucially, these distributions are more strongly peaked because of the non-linear response of the index strengths to the IMF slope. Expressed another way, the plateau in the X1 posterior distribution is due to the fact that the indices change little for variations in that range (and neither does f$_{\rm dwarf}$). It is therefore not immediately clear from the individual most probable values of X1 and X2 whether or not we find a bottom-heavy IMF, whereas examination of the joint distribution, or the f$_{\rm dwarf}$ distribution does indicate a modestly bottom-heavy IMF (though nevertheless consistent with Milky Way-like within the uncertainties). For this reason we ensure that all results presented in this work include f$_{\rm dwarf}$ and relative M/L values as well as the constraints on the power-law slopes (while cautioning the reader that the most probable values quoted may not appear to agree in an intuitive way).

\subsubsection*{Shape of the IMF}

\begin{figure*}
	\centering
	\includegraphics[width=0.99\textwidth]{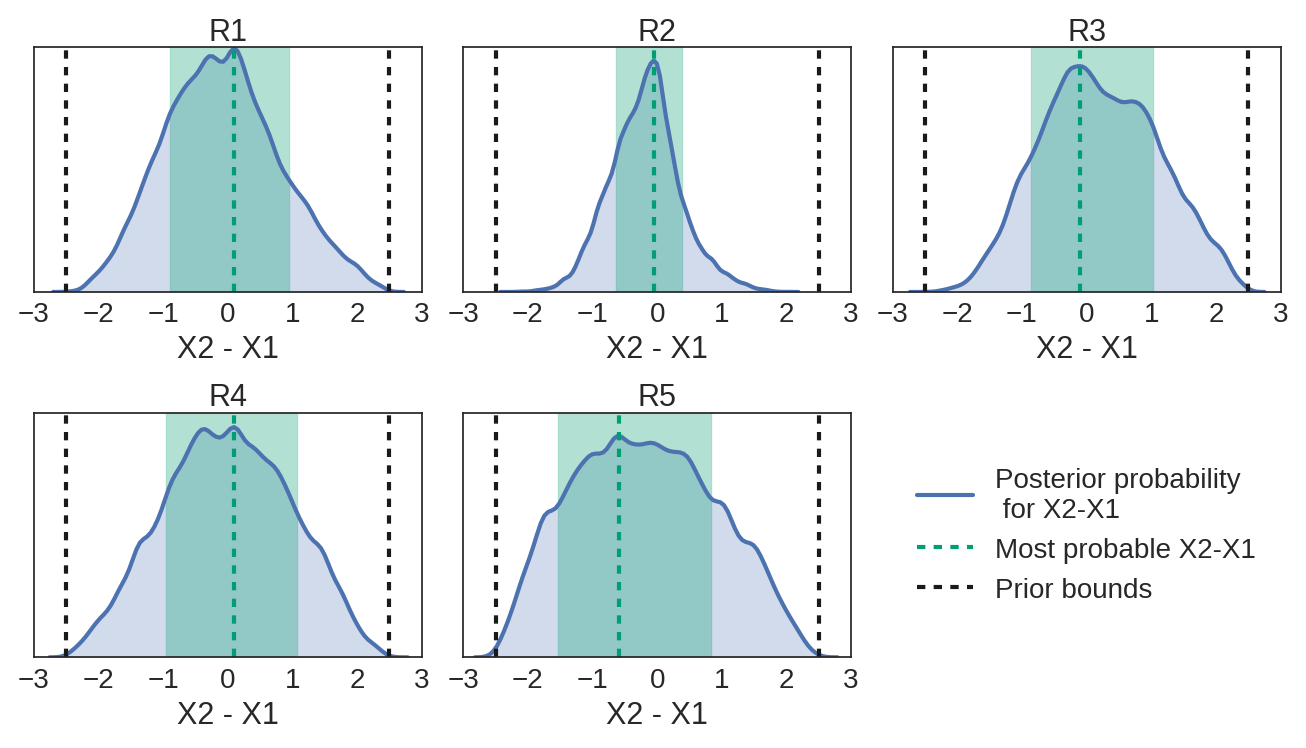}
	\caption[constraints on the IMF shape]{This plot indicates our constraints on the shape of the IMF via the difference in the power law slopes X1 and X2. The kernel density estimates show the posterior probability distributions of this quantity for the five stacked spectra. For the Milky Way IMF, X2--X1 = 1, whereas for a single power law (e.g. the Salpeter IMF) it is zero. The signal is weak since the IMF-dependent changes to spectral features principally track f$_{\rm dwarf}$. Because of this the posterior distributions reflect the imposed prior on X1 and X2, a point we explore in more detail in Appendix C.}
	\label{fig:shape_constraints}
\end{figure*}

Spectroscopic techniques for constraining the IMF are fairly insensitive to the functional form of the IMF, principally measuring the fractional contribution to the total light from dwarf stars. Nonetheless, given measurements of a wide set of absorption indices with sensitivities to different stellar mass ranges (as in this work), some information about the IMF shape can in principle be recovered. In practice, some covariance between the IMF broken power law slopes X1 and X2 remains, meaning the signal is weak. In Fig. \ref{fig:shape_constraints} we show the posterior probability distribution of the quantity X2--X1, a quantity which we use here as a proxy for IMF shape. The larger this quantity, the more sharply the IMF turns over at low stellar masses. Lines of constant X2--X1 lie approximately orthogonal to lines of constant f$_{\rm dwarf}$ in the (X1,X2) plane (e.g. Fig. \ref{fig:X1X2_joint_example}). For a Milky Way IMF X2--X1 = 1, while for a single power law such as the Salpeter IMF it is zero. While some studies have focussed on single power laws, others, for example \cite{2016MNRAS.457.1468L}, explore bottom-heavy IMFs with X1\,=\,1.3 (as in the Milky Way) and X2\,$>$\,2.3, i.e. X2--X1\,$>$\,1. For the stacked spectra, this proxy for IMF shape is not well-constrained, but appears to prefer (at a marginal level) IMFs that break less sharply at low mass. However, this is because with poor constraints on the IMF shape the marginalised distribution reflects the imposed prior. MCMC walkers exploring low f$_{\rm dwarf}$ regions are constrained more tightly in X2--X1 around zero, so if f$_{\rm dwarf}$ is not well-constrained in the middle of the prior box then when we marginalise over the full posterior distribution more walkers will be found in the vicinity of X2--X1\,=\,0 than if we had drawn uniformly from [f$_{\rm dwarf}$, X2--X1] instead. Note that in the work presented in Paper I we explicitly made use of a uniform prior in \fdwarf, while X2--X1 was implicitly fixed to zero.

\subsubsection*{Effect of the prior on inference of the IMF}

Evaluation of the posterior probability distribution for the IMF requires that a prior probability distribution be specified. In the work we have presented so far, we have assumed an equal prior probability for any value of X1 and X2 within set limits. This is not the only possible choice; an alternative would be to assume equal prior probability for any value of f$_{\rm dwarf}$ within certain limits and likewise for our proxy for IMF shape, X2--X1. Such a prior is very far from uniform in X1, X2.

\begin{figure*}
	\centering
	\includegraphics[width=0.68\textwidth]{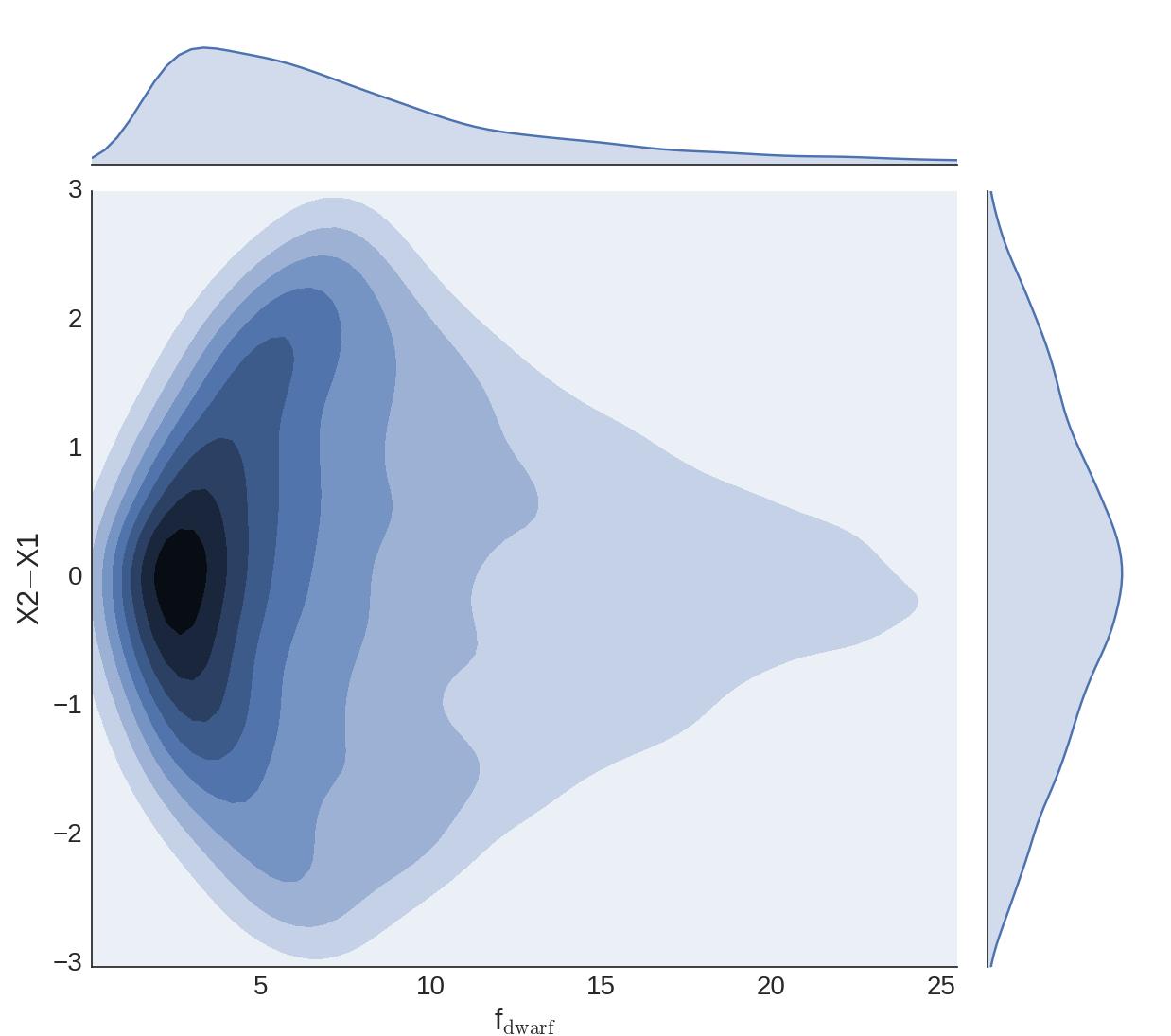}
	\caption[effect of prior probability distribution choice]{Density distribution in \fdwarf vs. X2--X1 space for points drawn uniformly from 0.5\,$<$\,X1, X2\,$<$\,3.5. This demonstrates that a prior probability distribution on X1 and X2 which is uniform between some limits (in this case, 0.5\,$<$\,X1, X2\,$<$\,3.5) corresponds to a peaked distribution in both \fdwarf and X2--X1.}
	\label{fig:prior_consequence}
\end{figure*}

In Fig. \ref{fig:prior_consequence} we demonstrate the difference between these two scenarios. The plot shows the density of points randomly and uniformly drawn from X1 and X2 within the bounds of the prior in \fdwarf, X2--X1 space as well as estimates of the marginalised distributions of these points, which we can identify with the corresponding prior on \fdwarf and X2--X1. It is clear that this prior picks out preferred values for these quantities. Likewise, a uniform prior on these quantities would pick out preferred values of X1, X2. Note that in the work presented in Paper I we explicitly made use of a uniform prior in \fdwarf.

\begin{figure*}
	\centering
	\includegraphics[width=0.99\textwidth]{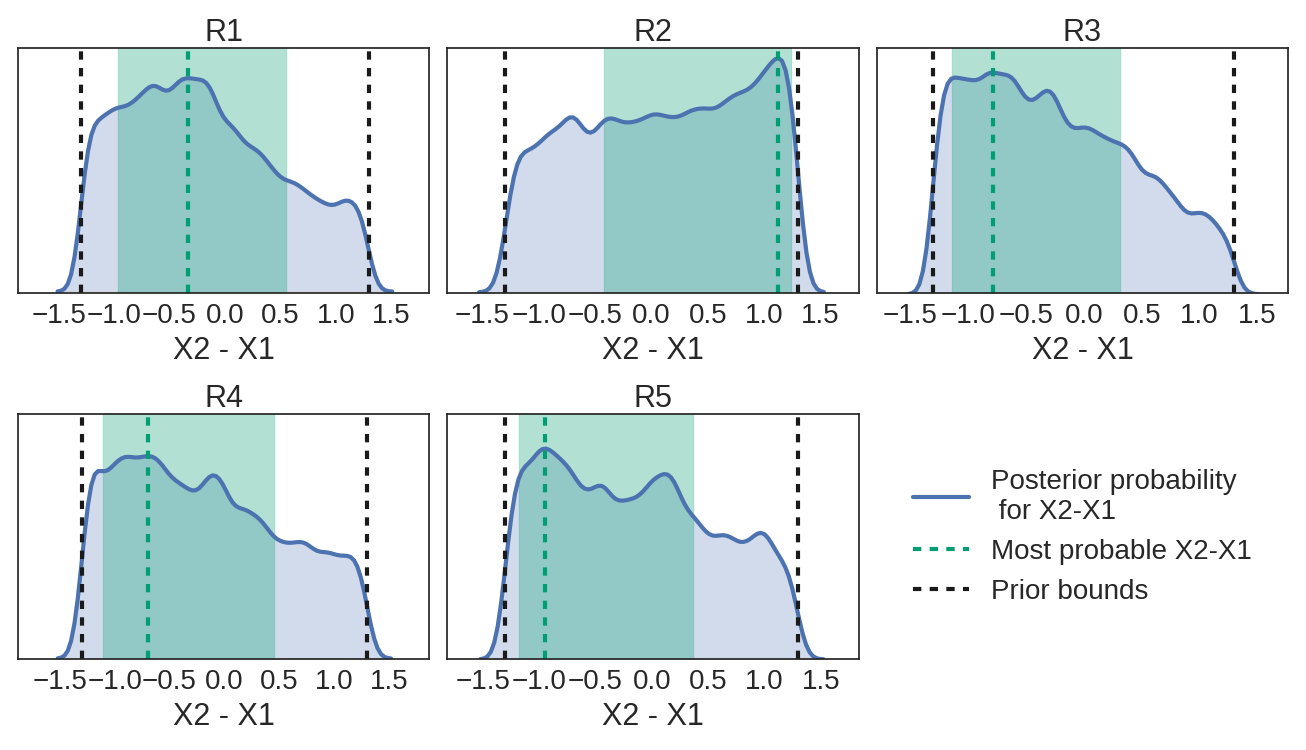}
	\caption[constraints on the IMF shape with alternative prior specification]{This plot indicates our constraints on the shape of the IMF via the difference in the power law slopes X1 and X2. The kernel density estimates show the posterior probability distributions of this quantity for the five stacked spectra. For the Milky Way IMF, X2--X1\,=\,1, whereas for a single power law (e.g. the Salpeter IMF) it is zero. The signal is weak since the IMF-dependent changes to spectral features principally track f$_{\rm dwarf}$. Because of this the posterior distributions reflect the imposed prior on X1 and X2; by contrast to Fig. \ref{fig:shape_constraints} a uniform prior on X2--X1 has been used here.}
	\label{fig:shape_constraints_UP}
\end{figure*}

It is crucial to quantify the effect that our choice of prior may have on our results, so to this end we re-ran MCMC using a uniform prior on both \fdwarf and X2--X1. It is important to note that this requires us to explore a more restricted parameter space (4\%\,$<$\,\fdwarf$<$15\%, --1.5\,$<$\,X2--X1\,$<$\,1.5) in order to remain within the CvD16 model grids. The results for most parameters are unchanged, but the prior unsurprisingly has an effect on the most probable values of the IMF parameters. In Fig. \ref{fig:shape_constraints_UP} we show the marginalised posterior distributions for X2--X1 computed for each of the stacked spectra. 

This makes clear that in some cases the evidence provided by our data (as averaged in the stacked spectra) generally favours IMFs that steepen at low stellar masses, but not decisively so. For Kroupa, X2--X1\,=\,1, which in the innermost stacked spectrum is somewhat in tension with the data (about 5\% of draws from the posterior distribution have X2--X1\,$>$\,1). On the other hand, whereas in Fig. \ref{fig:shape_constraints} the stack from the next extraction region out appeared to decisively exclude the Kroupa shape, Fig. \ref{fig:shape_constraints_UP} shows that the data are completely unable to constrain the shape under our alternative prior. It is clear that any attempt to extract information about the IMF shape from spectroscopic data must pay close attention to the chosen prior probability distribution.

In Table \ref{table:UP_results} we show how our results are modified using the alternative prior. This approach yields somewhat more bottom-heavy IMFs, which are nonetheless consistent with Kroupa (for which \fdwarf\,$\simeq$\,5\%). In the central extraction region the inferred \fdwarf and X2--X1 would also be consistent with a Salpeter single power-law, similar to the result obtained in Paper I (in which a prior that was uniform in \fdwarf was used). 

\begin{figure}
	\centering
	\includegraphics[width=0.49\textwidth]{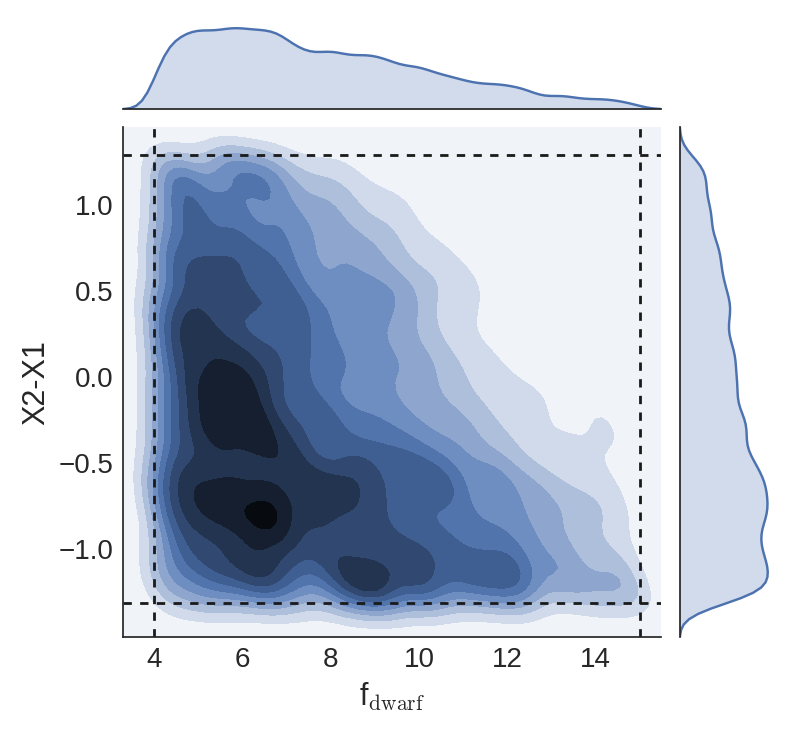}
	\caption[example joint posterior probability distribution over \fdwarf\, and IMF shape]{The results of fitting feature strengths from the centrally extracted spectrum from NGC\,5813 with a uniform prior on \fdwarf and X2--X1 (corresponds to Fig. \ref{fig:X1X2_joint_example}, other than the change to the prior). The joint and marginal distributions are shown via kernel-density estimation. Prior boundaries are indicated by the thick dashed lines. The results are consistent; a somewhat bottom-heavy IMF is recovered.}
	\label{fig:X1X2_joint_example_UP}
\end{figure}

Finally, in Fig. \ref{fig:X1X2_joint_example_UP} we show an equivalent plot to Fig. \ref{fig:X1X2_joint_example} but with the inference carried out under a prior that is uniform in \fdwarf and X2--X1. The results for this spectrum are comparable: a modestly bottom-heavy IMF is preferred, albeit one that steepens rather than flattens at lower stellar mass (note that the shapes are nonetheless consistent, given the uncertainties).

\begin{table*}
	\centering
	\caption[parameter estimates using alternative prior probability distribution]{Estimated parameter values for the stacked spectra, acquired using the updated models and the full set of index measurements. For the IMF we fit a two-part power law with slopes X1 and X2, but used a prior distribution that was uniform in \fdwarf and X2--X1 (instead of in X1 and X2). We also fit for stellar population age and total metallicity, as well as the abundances of various elements of interest (we note that the IMF prior has negligible impact on these).}
	\begin{tabular}{lcccccc}
		\hline
		&             R1             &            R2             &            R3             &            R4             &             R5             &      gradient      \\ \hline
		age /Gyr          & \gp{}10.75\rerr{1.43}{1.39} & \gp{}8.82\rerr{0.68}{0.92} & \gp{}8.47\rerr{0.90}{1.70} & \gp{}7.50\rerr{0.32}{3.33} & \gp{}8.86\rerr{1.22}{2.25}  &  --1.09$\pm$0.64   \\
		\fdwarf /\%       & \gp{}6.8\rerr{1.9}{3.4}  & \gp{}4.2\rerr{0.1}{0.9} & \gp{}4.9\rerr{0.6}{2.9} & \gp{}6.8\rerr{1.9}{4.0} & \gp{}10.0\rerr{3.8}{3.1} & \gp{}1.7$\pm$1.9 \\
		X2--X1            &   --0.33\rerr{0.63}{0.89}   & \gp{}1.13\rerr{1.55}{0.12} &  --0.79\rerr{0.35}{1.11}   &  --0.70\rerr{0.41}{1.16}   &   --0.95\rerr{0.23}{1.32}   &  --0.06$\pm$0.50   \\
		$\rm [Z/H]$       & \gp{}0.15\rerr{0.04}{0.04}  & \gp{}0.23\rerr{0.02}{0.02} & \gp{}0.13\rerr{0.06}{0.04} & \gp{}0.05\rerr{0.13}{0.06} &   --0.04\rerr{0.16}{0.08}   &  --0.06$\pm$0.04   \\
		$\rm [Fe/H]$      &   --0.18\rerr{0.10}{0.08}   &  --0.11\rerr{0.06}{0.04}   &  --0.19\rerr{0.13}{0.09}   &  --0.34\rerr{0.11}{0.16}   &   --0.21\rerr{0.20}{0.16}   &  --0.08$\pm$0.05   \\
		$\rm [Mg/Fe]$     & \gp{}0.57\rerr{0.08}{0.07}  & \gp{}0.45\rerr{0.04}{0.05} & \gp{}0.46\rerr{0.11}{0.07} & \gp{}0.50\rerr{0.14}{0.09} & \gp{}0.35\rerr{0.17}{0.18}  &  --0.05$\pm$0.04   \\
		$\rm [Na/Fe]$     & \gp{}0.84\rerr{0.13}{0.09}  & \gp{}0.71\rerr{0.05}{0.06} & \gp{}0.41\rerr{0.14}{0.14} & \gp{}0.25\rerr{0.31}{0.23} &   --0.17\rerr{0.26}{0.42}   &  --0.41$\pm$0.08   \\
		$\rm [Ca/Fe]$     & \gp{}0.24\rerr{0.07}{0.11}  & \gp{}0.11\rerr{0.05}{0.06} & \gp{}0.15\rerr{0.11}{0.09} & \gp{}0.15\rerr{0.16}{0.11} &   --0.08\rerr{0.17}{0.25}   &  --0.06$\pm$0.04   \\
		$\rm [O,\,Ne,\,S/Fe]$ & \gp{}0.83\rerr{0.22}{0.12}  & \gp{}0.67\rerr{0.09}{0.08} & \gp{}0.80\rerr{0.16}{0.12} & \gp{}0.76\rerr{0.26}{0.14} & \gp{}0.65\rerr{0.31}{0.23}  & \gp{}0.03$\pm$0.06 \\
		$\rm [C/Fe]$      & \gp{}0.22\rerr{0.11}{0.15}  & \gp{}0.13\rerr{0.05}{0.08} & \gp{}0.20\rerr{0.12}{0.13} & \gp{}0.27\rerr{0.17}{0.14} & \gp{}0.10\rerr{0.25}{0.19}  & \gp{}0.03$\pm$0.04 \\
		$\rm [Ti/Fe]$     & \gp{}0.61\rerr{0.15}{0.20}  & \gp{}0.32\rerr{0.11}{0.10} & \gp{}0.23\rerr{0.23}{0.20} & \gp{}0.53\rerr{0.27}{0.22} & \gp{}0.38\rerr{0.45}{0.35}  &  --0.07$\pm$0.09   \\
		$\rm [K/Fe]$      & \gp{}0.19\rerr{0.11}{0.19}  & \gp{}0.10\rerr{0.06}{0.09} & \gp{}0.20\rerr{0.15}{0.17} & \gp{}0.14\rerr{0.17}{0.25} & \gp{}0.28\rerr{0.38}{0.23}  & \gp{}0.05$\pm$0.08 \\ \hline
	\end{tabular}
	\label{table:UP_results}
\end{table*}

\section{Discussion}

We have measured the radially-varying strengths of several K-band absorption lines for seven of the eight ETGs in the  KINETyS sample, finding a significant gradient in the strength of the Na\,I 2.21\,$\upmu$m line and rather flat radial profiles for the Ca\,I features at 1.98\,$\upmu$m and 2.26\,$\upmu$m and also the CO bandhead at 2.30$\upmu$m.

Using the CvD16 SSP models and our own code for stellar population parameter fitting, we extended the work presented in Paper I. Our results are broadly consistent with those previously presented, but the more extensive parameter coverage of the new models -- combined with the additional data -- allow us to place better constraints on the IMF and chemical abundance properties of our sample than was previously possible.

As before, we do not find strong evidence for IMF variations in our sample of galaxies, either on average (via the stacked data) nor in individual galaxies. Indeed, the new parameter fits generally favour Milky Way-like IMFs and provide little evidence for an IMF gradient. Moreover, we now model the IMF as a two-part power law, mitigating any issues that might have arisen from our previous single parameter treatment, though at the expense of some additional ambiguity in the interpretation of the results. Our results contrast with e.g. \cite{2017ApJ...841...68V}, in which the IMF in was found to typically be very bottom-heavy (\fdwarf$>$20\%) in the core regions of ETGs and consistent with Milky Way-like at half the effective radius. On the other hand, they are not dissimilar from those in \cite{2017MNRAS.464.3597L}, in which two ETGs are studied and found (using an IMF model assumed to flatten below 0.5M$_\odot$) to have more modestly bottom-heavy IMFs in their core regions (\fdwarf$\gtrsim$8.4\%). Given the sensitivity of our results to the prior on the IMF this level of IMF variation is not necessarily inconsistent with our findings.

As stellar population synthesis models have become more sophisticated, especially with respect to their treatment of the IMF functional form, it has become apparent that IMF variations may be quite subtle in many cases. For example, \cite{2016arXiv161200065N} showed that a stacked spectrum derived from observations of high-mass ETGs ($\sigma \simeq 280$\,kms$^{-1}$) can, when modelled with a flexible IMF model consisting of a two-part power law and a variable low-mass cut-off, be reproduced with only a modest enhancement in stellar M/L (or \fdwarf). The same observations require a power law with a steeper-than-Salpeter slope if a single-slope IMF with fixed low-mass cut-off is assumed. Moreover, Newman et al. show that a model with such a flexible IMF can significantly reduce the tension between results from spectroscopy and from lensing measurements for the three lensed ETGs with Milky Way-like M/L presented in \cite{2015MNRAS.449.3441S}. If these fairly modest enhancements to the dwarf star content of ETGs are typical, the contribution of dwarf stars to the integrated light may be increased by only one or two percent in the majority of galaxies, which would be consistent with our findings here.

Concerning the recovered abundance gradients, in our new formalism we fit explicitly for total metallicity and then consider the variations of particular elements on top of this trend. On average, we find strong [Na/Fe] radial gradients in our sample, i.e. [Na/H] ramps up in the core faster than the overall metallicity trend would predict (by contrast, while Mg and perhaps Ca are enhanced, they are superabundant by a constant factor with respect to Fe throughout). This lends credence to our suggestion, first made in Paper I, that the metallicity-dependent yield of Na is playing a role in these galaxies. The extreme [Na/H] values we infer (typically 0.5--0.6 in the central extraction regions of the KINETyS sample, consistent with the result from the stacked spectrum for this region) potentially pose a challenge; SSP models must rely on theoretical predictions rather than empirical spectra since metal-rich stars so superabundant in Na are not observed in the Milky Way.

Our results indicate some tension between the measurements of different Na lines, which might be accounted for by a modelling issue of this kind. In Fig. \ref{fig:na_idx_idx} we showed how for the stacked spectra the Na\,I 0.82\,$\upmu$m line strength at face value appears to prefer lower values of [Na/Fe]. Index vs. index grids should be treated with caution due to the cumulative minor effects of various unaccounted-for parameters. For example, in this case the inclusion of a [Ti/Fe] parameter (as in the more complex model presented above) can reduce the tension significantly, though not entirely. This is because Na\,I 0.82\,$\upmu$m lies within a broad Ti absorption band. The effect of this enhanced Ti is to weaken the measured equivalent-width somewhat. This and other parameters can therefore help to reconcile the measurements by allowing weaker Na\,I 0.82\,$\upmu$m at higher [Na/Fe]. Nevertheless, some systematic tension remains between the three infrared Na\,I features, which may be partially due to the modelling of these features in highly Na-enhanced populations (but see \citealp{2017MNRAS.464.3597L}).

\begin{figure}
	\centering
	\includegraphics[width=0.49\textwidth]{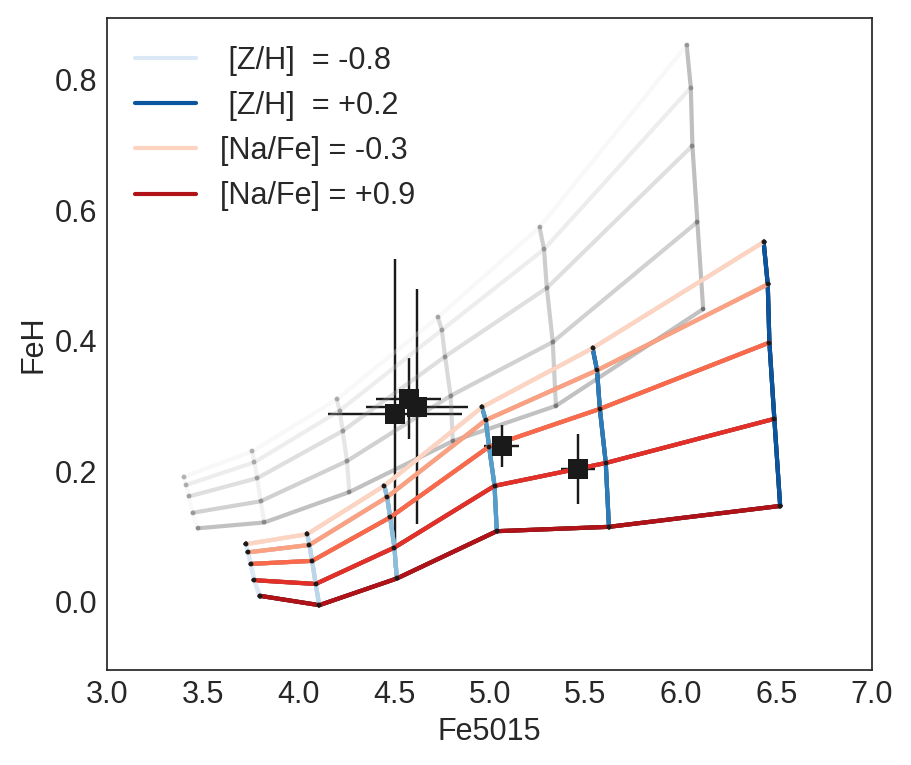}
	\caption[index-index plot for Fe\,5015 and the Wing-Ford (FeH) band]{Index-index grid for the optical Fe\,5015 index and the Wing-Ford (FeH) band, computed at [$\alpha$/Fe]\,=\,+0.5 for a Kroupa IMF. The KMOS data are superimposed in black. The stacked measurements of the Wing-Ford band are weakest for the innermost extraction region (e.g. see Fig. \ref{fig:model_index_predictions}), consistent with the strongly increasing Na abundance, which barely affects Fe\,5015 but has a significant effect on the Wing-Ford band. Note that a more bottom-heavy IMF would tend to make FeH stronger while leaving Fe\,5015 largely unchanged -- the shift for a Salpeter single power law is indicated by the offset light grey grid -- so would require e.g. even more extreme Na abundance to match the KMOS measurements.}
	\label{fig:fe_feh_idx}
\end{figure}

Proper constraints on [Na/Fe] are crucial, as Na abundance has important effects on the strength of other indices, in particular the Wing-Ford band. In Fig. \ref{fig:fe_feh_idx} we show how this effect can explain the radial behaviour of the Wing-Ford band in the stacked KMOS spectra. The same figure also shows that it would be challenging to reconcile this measurement with a systematically bottom-heavy IMF in the cores of our sample galaxies, in contrast to the results of e.g. \cite{2015MNRAS.447.1033M} and \cite{2017ApJ...841...68V}. Note, however, that an even more flexible IMF prescription that included a flexible low-mass cutoff in the IMF could help mitigate this tension, since the Wing-Ford band is principally sensitive to the lowest mass dwarf stars.

The Wing-Ford band in massive ETGs has undergone extensive study in recent years, e.g. by \cite{2016ApJ...821...39M}, \cite{2016arXiv161200364V}, \cite{2017MNRAS.465..192Z}, and \cite{2016MNRAS.457.1468L}. These studies find a range of values for the Wing-Ford (FeH) band, from 0.4--0.6\AA, whereas our measurements from KMOS data lie primarily between 0.2--0.3\AA. However, it is notable that e.g. Zieleniewski quantifies possible systematics affecting FeH and concludes that true values of the strength as low as 0.2--0.3\AA\, (at 200\,kms$^{-1}$ velocity dispersion) are not implausible; FeH is a weak feature and so is hard to measure accurately. Therefore this tension may simply reflect the varying systematics affecting different studies. In the case of our results, we do not find that removing FeH from consideration leads to more bottom-heavy IMFs being recovered (although of course the statistical uncertainty is on the value of \fdwarf is somewhat increased -- by 6\% on average in the stacked spectra -- without it).

\section{Conclusions}

We have built upon the work presented in Paper I in several important ways. First, we have deployed a significantly more advanced implementation of our stellar population parameter inference code. The updated code capitilises on recent advances by using the state-of-the-art CvD16 stellar population models. These are based on a much-improved stellar spectral library and feature a more comprehensive set of stellar population parameters than the CvD12 models that we used previously, including the ability to model IMFs with different shapes and to vary the overall metallicity. Secondly, we have obtained K-band spectroscopy of almost all of the KINETyS sample, allowing us to make measurements of previously underutilised spectral features as stellar population diagnostics. This adds significant discriminatory power to our method, allowing us to further break degeneracies between model parameters. Our conclusions are as follows:

\begin{enumerate}
	
	\item We used stacked spectra extracted from different radii within the KINETyS sample to investigate the radial behaviour of four K-band spectral features. We found a strong decline in the strength of the Na\,I 2.21$\upmu$m line with radius ($-0.53\pm0.08$ per decade in R/R$_{\rm eff}$), and fairly constant strength for two Ca\,I features. The CO 2.30\umicron\,$a$ line likewise appears roughly constant in strength ($-0.11\pm0.13$).
	\item We use our updated stellar population spectral index fitting code in conjunction with these measurements to place additional constraints on the stellar populations of our sample. From the stacked spectra we recover an average metallicity gradient $\Delta$[Z/H] of $-0.11\pm0.03$ per decade in R/R$_{\rm eff}$ and measure the abundances of several $\alpha$-capture elements, finding flat trends in the relative abundances [Mg/Fe], [Ca/Fe], [Ti/Fe], and [O,\,Ne,\,S/Fe]. The latter two parameters are not well constrained, but we find [Ca/Fe] much less enhanced than [Mg/Fe], consistent with our previous findings.
	\item We find a very strong radial gradient in [Na/Fe] of $-0.35\pm0.09$, a tighter constraint than that presented in Paper I, aided by our measurements of the Na\,I 2.21$\upmu$m line. However, there is mild tension between the absolute strengths of the three Na\,I features we measure, which may be a consequence of the theoretical modelling of Na line strengths in stellar populations where Na is superabundant. The extreme abundance of Na implied by our measurements in the cores of massive ETGs may be explained by the metallicity-dependent nucleosynthetic yield of Na.
	\item For the IMF, we infer a Milky Way-like, non-radially varying IMF both for the stacked spectra and most of the individual galaxies in the sample.
	
\end{enumerate}
	
Our results appear to indicate that the principal driver of radial spectroscopic variations is the abundance gradients created by the ETG formation process. On average, we infer that IMF variations in massive ETGs are subtle in most cases and we do not find compelling evidence for strong radial variations in the IMF in the inner regions. This stands in contrast to some recent results. In future it will be important to compare constraints from different methods (e.g. spectroscopy and dynamics) in individual galaxies in order to address this tension.
	
\section*{Acknowledgements}
	
PA was supported by an STFC studentship (ST/K501979/1) and RJS by the STFC Durham Astronomy Consolidated Grants 2014--2017 (ST/L00075X/1) and 2017--2020 (ST/P000541/1). The KMOS data used here are available through the ESO science archive (programme ID: 097.B-0882(A)).

The authors thank Charlie Conroy for providing the latest version of the CvD models.
	
	
	
	
	
\bibliographystyle{mnras}
\bibliography{PAlton} 

	

\clearpage

	\appendix
	
\section{Measurements of absorption feature indices}
	
The full set of measurements for individual galaxies and the stacked spectra are presented in this appendix. These measurements are corrected to 230\,kms$^{-1}$. The velocity dispersions of the K-band stacked spectra were measured using pPXF \citep{2004PASP..116..138C} using the spectral region around the CO bandhead (21200--23400\AA). These velocity dispersion measurements are also given in the data tables.

The measured equivalent widths for the KMOS data are presented galaxy-by-galaxy in Tables \ref{table:t1407} to \ref{table:t5813}. The data from the stacked spectra are given in Table \ref{table:kmos_k_stacks}.

\begin{table*}
	\centering
	\caption[K-band data for NGC\,1407]{A list of index measurements (in \AA) for the K-band observations of NGC\,1407, corrected to 230\,kms$^{-1}$. R1 and R2 correspond to the two extraction regions in the central IFU ($<0.7\arcsec$, $>0.7\arcsec$) while R3, R4, and R5 correspond to the rings of IFUs arranged at $\sim1/3\,$R$_{\rm eff}$, $\sim2/3\,$R$_{\rm eff}$, and $\sim$R$_{\rm eff}$.}
	\label{table:t1407}
	\begin{tabular}{c|ccccc}
		\hline
		NGC1407       &      R1       &      R2       &         R3         &          R4          &          R5          \\ \hline
		Ca\,I 1.98\umicron  & 0.61$\pm$4.15 & 2.53$\pm$5.05 &  --6.46$\pm$9.09   &  --21.67$\pm$28.53   & \gp{}29.21$\pm$69.27 \\
		Na\,I 2.21\umicron  & 2.75$\pm$0.32 & 2.85$\pm$0.98 & \gp{}1.65$\pm$0.49 &  \gp{}2.39$\pm$1.65  &   --3.21$\pm$2.74    \\
		Ca\,I 2.26\umicron  & 3.17$\pm$0.65 & 3.10$\pm$1.14 & \gp{}3.67$\pm$2.49 &  \gp{}1.84$\pm$7.82  &  \gp{}3.25$\pm$5.98  \\
		CO 2.30\umicron\, a & 6.65$\pm$0.49 & 6.82$\pm$0.52 & \gp{}6.95$\pm$1.88 &  \gp{}9.63$\pm$9.40  &  \gp{}4.91$\pm$4.61  \\
		CO 2.30\umicron\, b & 4.36$\pm$1.36 & 8.02$\pm$1.44 & \gp{}8.03$\pm$5.69 & \gp{}23.71$\pm$45.27 & \gp{}6.31$\pm$12.01  \\ \hline
		average S/N     &      71       &      24       &         33         &          14          &          12          \\ \hline
	\end{tabular}
\end{table*}

\begin{table*}
	\centering
	\caption[K-band data for NGC\,3377]{A list of index measurements (in \AA) for the K-band observations of NGC\,3377, corrected to 230\,kms$^{-1}$. R1 and R2 correspond to the two extraction regions in the central IFU ($<0.7\arcsec$, $>0.7\arcsec$) while R3, R4, and R5 correspond to the rings of IFUs arranged at $\sim1/3\,$R$_{\rm eff}$, $\sim2/3\,$R$_{\rm eff}$, and $\sim$R$_{\rm eff}$.}
	\label{table:t3377}
	\begin{tabular}{c|ccccc}
		\hline
		NGC3377 & R1 & R2 & R3 & R4 & R5 \\ \hline
		Ca\,I 1.98\umicron & 1.13$\pm$0.15 & 0.80$\pm$0.20 & 0.90$\pm$0.35 & 1.42$\pm$0.77 & 0.55$\pm$1.11 \\
		Na\,I 2.21\umicron & 1.73$\pm$0.11 & 1.34$\pm$0.22 & 1.00$\pm$0.45 & 1.89$\pm$0.88 & 3.20$\pm$1.39 \\
		Ca\,I 2.26\umicron & 3.60$\pm$0.32 & 2.41$\pm$0.48 & 3.18$\pm$1.65 & 8.02$\pm$2.32 & 2.37$\pm$4.13 \\
		CO 2.30\umicron\, a & 8.33$\pm$0.18 & 7.96$\pm$0.21 & 5.52$\pm$1.73 & 6.23$\pm$2.75 & 2.67$\pm$5.15 \\
		CO 2.30\umicron\, b & 6.94$\pm$0.20 & 6.58$\pm$0.22 & 5.58$\pm$2.17 & 376.21$\pm$335.59 & 1.32$\pm$5.59 \\
		\hline
		average S/N & 71 & 24 & 33 & 14 & 12 \\
		\hline
	\end{tabular}
\end{table*}

\begin{table*}
	\centering
	\caption[K-band data for NGC\,3379]{A list of index measurements (in \AA) for the K-band observations of NGC\,3379, corrected to 230\,kms$^{-1}$. R1 and R2 correspond to the two extraction regions in the central IFU ($<0.7\arcsec$, $>0.7\arcsec$) while R3, R4, and R5 correspond to the rings of IFUs arranged at $\sim1/3\,$R$_{\rm eff}$, $\sim2/3\,$R$_{\rm eff}$, and $\sim$R$_{\rm eff}$.}
	\label{table:t3379}
	\begin{tabular}{c|ccccc}
		\hline
		NGC3379 & R1 & R2 & R3 & R4 & R5 \\ \hline
		Ca\,I 1.98\umicron & 1.37$\pm$0.22 & 1.31$\pm$0.84 & 0.96$\pm$0.10 & 0.70$\pm$0.15 & 0.81$\pm$0.23 \\
		Na\,I 2.21\umicron & 2.26$\pm$0.18 & 2.01$\pm$0.93 & 1.67$\pm$0.20 & 1.39$\pm$0.44 & 1.39$\pm$0.78 \\
		Ca\,I 2.26\umicron & 3.60$\pm$0.29 & 3.61$\pm$0.77 & 3.80$\pm$0.46 & 3.26$\pm$1.04 & 2.71$\pm$1.83 \\
		CO 2.30\umicron\, a & 8.73$\pm$0.10 & 8.38$\pm$0.20 & 7.29$\pm$0.41 & 7.07$\pm$1.04 & 8.40$\pm$1.54 \\
		CO 2.30\umicron\, b & 6.24$\pm$0.15 & 5.91$\pm$0.27 & 5.29$\pm$0.58 & 4.79$\pm$1.66 & 5.47$\pm$2.69 \\
		\hline
		average S/N & 71 & 24 & 33 & 14 & 12 \\
		\hline
	\end{tabular}
\end{table*}

\begin{table*}
	\centering
	\caption[K-band data for NGC\,4486]{A list of index measurements (in \AA) for the K-band observations of NGC\,4486, corrected to 230\,kms$^{-1}$. R1 and R2 correspond to the two extraction regions in the central IFU ($<0.7\arcsec$, $>0.7\arcsec$) while R3, R4, and R5 correspond to the rings of IFUs arranged at $\sim1/3\,$R$_{\rm eff}$, $\sim2/3\,$R$_{\rm eff}$, and $\sim$R$_{\rm eff}$.}
	\label{table:t4486}
	\begin{tabular}{c|ccccc}
		\hline
		NGC4486 & R1 & R2 & R3 & R4 & R5 \\ \hline
		Ca\,I 1.98\umicron & 2.17$\pm$2.97 & 1.74$\pm$0.22 & 2.09$\pm$0.27 & 1.23$\pm$0.40 & 1.48$\pm$0.88 \\
		Na\,I 2.21\umicron & 1.83$\pm$1.33 & 1.85$\pm$0.19 & 2.48$\pm$0.32 & 1.16$\pm$0.66 & 1.46$\pm$0.98 \\
		Ca\,I 2.26\umicron & 2.10$\pm$0.79 & 3.93$\pm$0.44 & 3.97$\pm$1.01 & 2.79$\pm$1.91 & 2.32$\pm$3.34 \\
		CO 2.30\umicron\, a & 5.20$\pm$0.42 & 7.47$\pm$0.30 & 9.16$\pm$0.90 & 8.40$\pm$1.77 & 8.54$\pm$3.20 \\
		CO 2.30\umicron\, b & 4.42$\pm$0.43 & 5.82$\pm$0.37 & 7.53$\pm$1.11 & 5.01$\pm$2.06 & 7.50$\pm$3.44 \\
		\hline
		average S/N & 71 & 24 & 33 & 14 & 12 \\
		\hline
	\end{tabular}
\end{table*}

\begin{table*}
	\centering
	\caption[K-band data for NGC\,4552]{A list of index measurements (in \AA) for the K-band observations of NGC\,4552, corrected to 230\,kms$^{-1}$. R1 and R2 correspond to the two extraction regions in the central IFU ($<0.7\arcsec$, $>0.7\arcsec$) while R3, R4, and R5 correspond to the rings of IFUs arranged at $\sim1/3\,$R$_{\rm eff}$, $\sim2/3\,$R$_{\rm eff}$, and $\sim$R$_{\rm eff}$.}
	\label{table:t4552}
	\begin{tabular}{c|ccccc}
		\hline
		NGC4552 & R1 & R2 & R3 & R4 & R5 \\ \hline
		Ca\,I 1.98\umicron & 2.96$\pm$0.27 & 3.01$\pm$0.29 & 1.79$\pm$0.28 & 0.23$\pm$0.64 & 0.59$\pm$0.70 \\
		Na\,I 2.21\umicron & 2.29$\pm$0.21 & 2.15$\pm$0.25 & 1.22$\pm$0.37 & 1.72$\pm$0.85 & 1.93$\pm$1.03 \\
		Ca\,I 2.26\umicron & 3.44$\pm$0.27 & 3.26$\pm$0.26 & 2.34$\pm$0.98 & 3.31$\pm$2.01 & 3.59$\pm$2.63 \\
		CO 2.30\umicron\, a & 8.39$\pm$0.33 & 8.14$\pm$0.31 & 6.61$\pm$0.85 & 6.58$\pm$1.77 & 8.15$\pm$2.50 \\
		CO 2.30\umicron\, b & 6.55$\pm$0.16 & 6.71$\pm$0.15 & 5.78$\pm$0.94 & 7.16$\pm$2.12 & 6.11$\pm$2.96 \\
		\hline
		average S/N & 71 & 24 & 33 & 14 & 12 \\
		\hline
	\end{tabular}
\end{table*}

\begin{table*}
	\centering
	\caption[K-band data for NGC\,4621]{A list of index measurements (in \AA) for the K-band observations of NGC\,4621, corrected to 230\,kms$^{-1}$. R1 and R2 correspond to the two extraction regions in the central IFU ($<0.7\arcsec$, $>0.7\arcsec$) while R3, R4, and R5 correspond to the rings of IFUs arranged at $\sim1/3\,$R$_{\rm eff}$, $\sim2/3\,$R$_{\rm eff}$, and $\sim$R$_{\rm eff}$.}
	\label{table:t4621}
	\begin{tabular}{c|ccccc}
		\hline
		NGC4621 & R1 & R2 & R3 & R4 & R5 \\ \hline
		Ca\,I 1.98\umicron & 0.16$\pm$0.23 & 0.44$\pm$0.52 & 0.09$\pm$0.22 & 0.54$\pm$0.40 & 1.34$\pm$0.48 \\
		Na\,I 2.21\umicron & 3.31$\pm$0.28 & 2.73$\pm$0.53 & 2.18$\pm$0.52 & 0.66$\pm$1.06 & 1.77$\pm$1.67 \\
		Ca\,I 2.26\umicron & 2.69$\pm$0.26 & 2.15$\pm$0.59 & 1.83$\pm$1.32 & 1.62$\pm$2.56 & 2.66$\pm$4.87 \\
		CO 2.30\umicron\, a & 8.57$\pm$0.41 & 8.35$\pm$0.47 & 7.62$\pm$1.30 & 8.26$\pm$3.05 & 105.53$\pm$33.77 \\
		CO 2.30\umicron\, b & 7.64$\pm$0.24 & 7.06$\pm$0.35 & 6.70$\pm$1.81 & 6.34$\pm$4.95 & 7.12$\pm$73.43 \\
		\hline
		average S/N & 71 & 24 & 33 & 14 & 12 \\
		\hline
	\end{tabular}
\end{table*}

\begin{table*}
	\centering
	\caption[K-band data for NGC\,5813]{A list of index measurements (in \AA) for the K-band observations of NGC\,5813, corrected to 230\,kms$^{-1}$. R1 and R2 correspond to the two extraction regions in the central IFU ($<0.7\arcsec$, $>0.7\arcsec$) while R3, R4, and R5 correspond to the rings of IFUs arranged at $\sim1/3\,$R$_{\rm eff}$, $\sim2/3\,$R$_{\rm eff}$, and $\sim$R$_{\rm eff}$.}
	\label{table:t5813}
	\begin{tabular}{c|ccccc}
		\hline
		NGC5813 & R1 & R2 & R3 & R4 & R5 \\ \hline
		Ca\,I 1.98\umicron & 1.86$\pm$0.65 & 2.27$\pm$0.78 & 3.86$\pm$3.59 & 5.98$\pm$4.70 & 1.29$\pm$12.11 \\
		Na\,I 2.21\umicron & 1.94$\pm$0.27 & 1.97$\pm$0.65 & 1.71$\pm$1.32 & 1.05$\pm$1.98 & 2.98$\pm$4.11 \\
		Ca\,I 2.26\umicron & 3.40$\pm$0.41 & 3.35$\pm$0.72 & 1.19$\pm$3.34 & 3.67$\pm$5.30 & 3.60$\pm$8.98 \\
		CO 2.30\umicron\, a & 6.84$\pm$0.39 & 7.02$\pm$0.46 & 7.64$\pm$2.94 & 8.43$\pm$5.21 & 0.40$\pm$10.77 \\
		CO 2.30\umicron\, b & 6.41$\pm$1.31 & 6.18$\pm$1.34 & 4.58$\pm$10.22 & 7.50$\pm$14.47 & 15.56$\pm$36.31 \\
		\hline
		average S/N & 71 & 24 & 33 & 14 & 12 \\
		\hline
	\end{tabular}
\end{table*}

\begin{table*}
	\centering
	\caption[K-band data for KMOS stacked spectra]{A list of index measurements (in \AA) for the stacked spectra, corrected to 230\,kms$^{-1}$. N.B. `noise' in S/N calculation represents scatter between input spectra, not the formal flux uncertainty (which is much lower). This S/N distribution is strongly skewed (e.g. 95\% range of central stack is 20---1780). R1 and R2 correspond to the two extraction regions in the central IFU ($<0.7\arcsec$, $>0.7\arcsec$) while R3, R4, and R5 correspond to the rings of IFUs arranged at $\sim1/3\,$R$_{\rm eff}$, $\sim2/3\,$R$_{\rm eff}$, and $\sim$R$_{\rm eff}$.}
	\begin{tabular}{lccccc}
		\hline
		Stacks                           &       R1       &       R2       &       R3       &       R4       &       R5       \\ \hline
		Ca\,I 1.98\umicron                & 0.44$\pm$0.22  & 0.40$\pm$0.12  & 0.48$\pm$0.12  & 0.53$\pm$0.21  & 0.37$\pm$0.30  \\
		Na\,I 2.21\umicron                & 1.82$\pm$0.14  & 1.98$\pm$0.05  & 1.63$\pm$0.16  & 0.96$\pm$0.24  & 0.94$\pm$0.30  \\
		Ca\,I 2.26\umicron                & 2.94$\pm$0.11  & 2.87$\pm$0.07  & 2.74$\pm$0.16  & 3.12$\pm$0.33  & 2.73$\pm$0.45  \\
		CO 2.30\umicron\, a               & 7.14$\pm$0.15  & 7.45$\pm$0.06  & 7.27$\pm$0.21  & 7.13$\pm$0.33  & 5.99$\pm$0.64  \\
		CO 2.30\umicron\, b               & 5.45$\pm$0.21  & 5.98$\pm$0.05  & 6.34$\pm$0.24  & 5.56$\pm$0.48  & 6.42$\pm$0.61  \\ \hline
		$\sigma_{\rm stack}$ /kms$^{-1}$ &   199$\pm$2    &   211$\pm$1    &   192$\pm$3    &   166$\pm$10   &   151$\pm$14   \\
		S/N (68\% range)                 &    70---600    &   170---1210   &    60---490    &    30---250    &    20---160    \\ \hline
		\label{table:kmos_k_stacks}      &
	\end{tabular}
\end{table*}

\section{Inference of the IMF shape}

It is possible to characterise the IMF in a variety of ways. In Paper I, we chose to use the quantity \fdwarf, the fractional contribution of dwarf stars to the total light, to parametrize the IMF. This choice was motivated by the linearity of the response of the model equivalent widths to changes in \fdwarf, a property not shared by other possible parametrizations such as the fractional contribution of dwarf stars to the total stellar mass, or indeed the IMF power law slope. Moreover, we found that for the spectral features analysed in Paper I the strength of IMF-sensitive features was approximately independent of the particular shape of the IMF. 

\begin{figure*}
	\centering
	\includegraphics[width=0.98\textwidth]{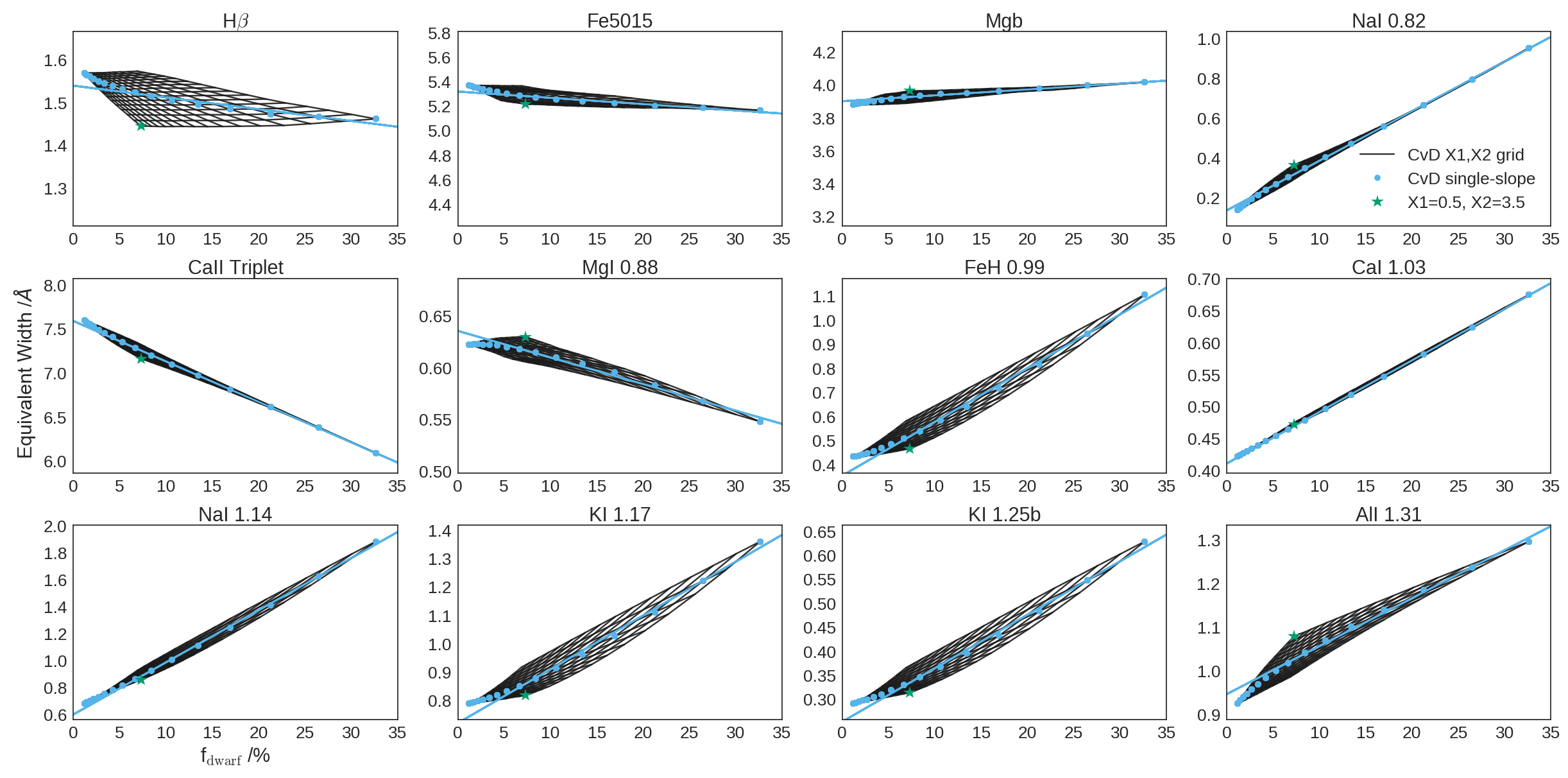}
	\caption[index vs. \fdwarf\, grids for a flexible IMF]{A set of spectral indices, computed for a flexible two-part power-law IMF (described in the text) with slope X1 below 0.5\,M$_\odot$ and X2 above. Units are index equivalent width in \AA. The black grid shows the predictions of the model assuming various combinations of values for X1 and X2 between 0.5 and 3.5. The blue points mark the diagonals of the grid (X1\,=\,X2, the single power-law case) and the blue line is a linear fit to these points for f$_{\rm dwarf}>$5\%. In each panel the most convex IMF is marked with a green star.}
	\label{fig:fdwarf_bpl}
\end{figure*}

We make a clearer demonstration of this point in Fig. \ref{fig:fdwarf_bpl}. For each of the spectral indices considered in Paper I we show \fdwarf\, plotted against index strength for a grid of IMFs parametrized according to a two-part broken power law, comprising two sections (0.08\,M$_{\odot}$\,--\,0.5\,M$_{\odot}$ and 0.5\,M$_{\odot}$\,--\,1.0\,M$_{\odot}$) with independent power-law slopes X1 and X2. For this grid X1, X2 run from 0.5 to 3.5 (where X1\,=\,X2\,=\,2.3 corresponds to the Salpeter IMF).

\begin{figure*}
	\centering
	\includegraphics[width=0.98\textwidth]{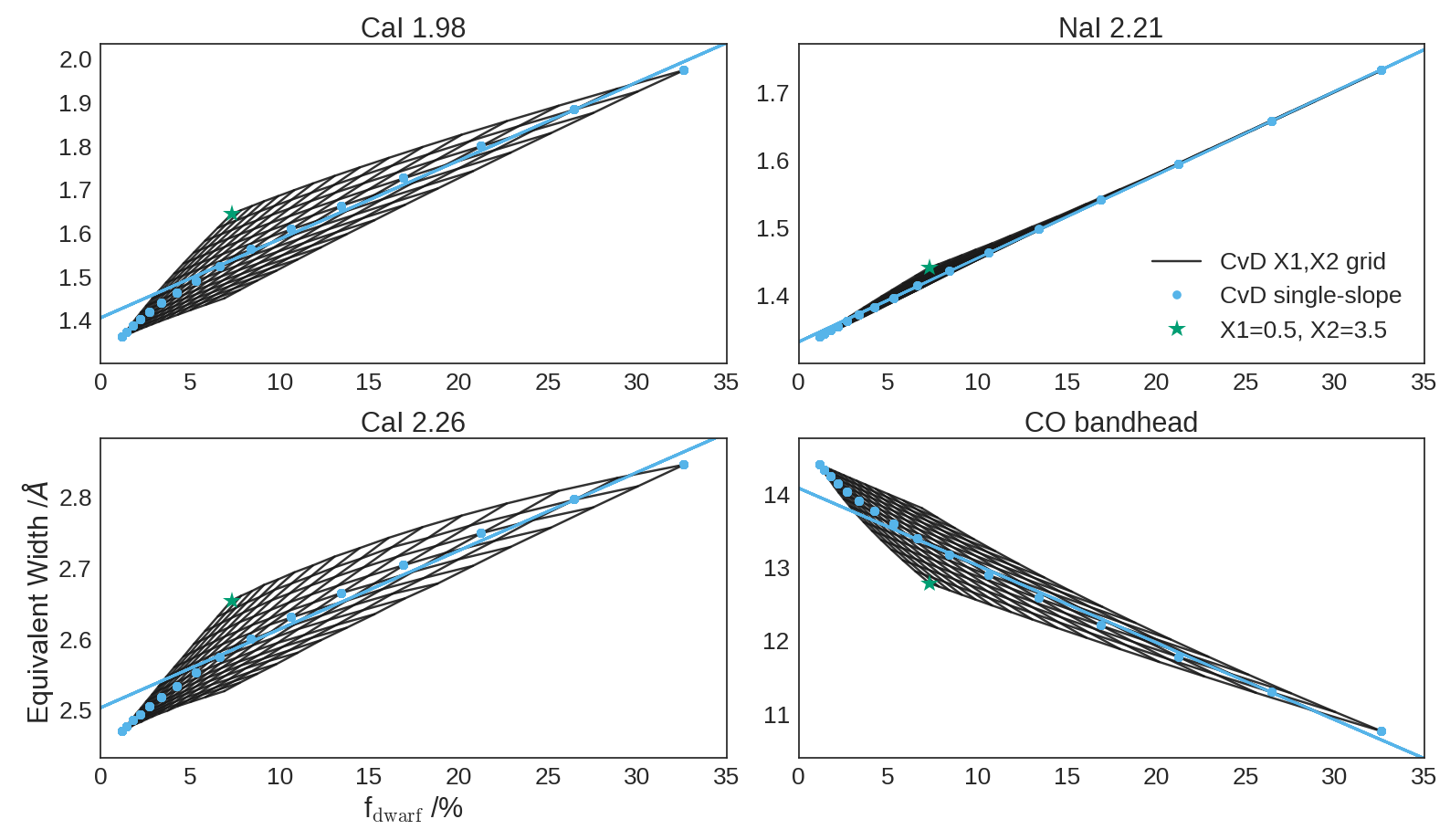}
	\caption[K-band index vs. \fdwarf\, grids for a flexible IMF]{Four K-band spectral indices, computed for a flexible two-part power-law IMF (described in the text). Units are index equivalent widths in \AA. In this case some indices have non-negligible dependence on the detailed IMF shape: the one-to-one relationship between f$_{\rm dwarf}$ and feature strength given by the blue line only holds strongly for Na\,I 2.21\umicron. In each panel the most convex IMF is marked with a green star.}
	\label{fig:fdwarf_bpl_K}
\end{figure*}

Fig. \ref{fig:fdwarf_bpl} shows that in general the grids fold up tightly around the linear relation we assumed in Paper I (e.g. Na\,I 0.82\umicron, Na\,I 1.14\umicron, Ca\,II Triplet). Tthe parameters X1 and X2 have some degeneracy and feature strengths principally track \fdwarf. Nevertheless, shape-dependent deviations clearly do matter in some cases (e.g. FeH 0.99\umicron\, and H$\beta$ strength can both vary by $\lesssim$10\% at fixed \fdwarf). It is also noteworthy that for bottom-light IMFs (lower fraction of dwarf stars than the Milky Way case, which can be approximated as X1\,=\,1.3, X2\,=\,2.3) the linear relation tends not to be a good model. However, in different wavelength regimes, different stars dominate the total luminosity. One way to extract more information about the detailed shape of the IMF may be to consider spectral indices over a long wavelength range. In Fig. \ref{fig:fdwarf_bpl_K} a similar set of relationships to those depicted in Fig. \ref{fig:fdwarf_bpl} are shown for four K-band ($\sim$\,2\umicron) spectral features; all of these have significant sensitivity to the IMF and all apart from Na\,I 2.21\umicron\, have a fairly strong dependence on the IMF shape.

This suggests that with a sufficiently wide range of spectral feature measurements, we may be able to recover some information about the shape of the IMF, or else at the least should marginalise over reasonable variations in shape to ensure that we obtain robust results. We therefore model the measured features with a broken power law as just described, calculating \fdwarf\, \textit{post hoc}. When necessary to discuss our inference of the IMF shape, we use the quantity X2--X1, which (as described in Section 6.3) varies approximately independently of \fdwarf in the CvD16 models.


	\bsp	
	\label{lastpage}
\end{document}